\documentclass[%
 reprint,
amsmath,amssymb,
pra
]{revtex4-2}

\usepackage{graphicx}
\usepackage{dcolumn}
\usepackage{bm}
\usepackage{amsmath}
\usepackage{color}
\usepackage{hyperref}
\hypersetup{
   colorlinks=true,
    linkcolor=blue,
    filecolor=blue,      
    urlcolor=blue,
    citecolor=cyan,
}

\begin{document}

\preprint{APS/123-QED}

\title{Experimental 3D super-localization with Laguerre-Gaussian modes}

\author{Chenyu Hu$^{1}$, Liang Xu$^{2}$, Ben Wang$^{1}$, Zhiwen Li $^{1}$, Yipeng Zhang $^{1}$}
\author{Yong Zhang$^{1}$}
\author{Lijian Zhang$^{1}$}\email{lijian.zhang@nju.edu.cn}

\affiliation{$^{1}$National Laboratory of Solid State Microstructures,
Key Laboratory of Intelligent Optical Sensing and Manipulation, College of Engineering and Applied Sciences, School of Physics, and Collaborative Innovation Center of Advanced Microstructures, Nanjing University, Nanjing 210093, China\\$^{2}$Research Center for Quantum Sensing, Zhejiang Lab, Hangzhou 310000, China}


\begin{abstract}
Improving three-dimensional (3D) localization precision is of paramount importance for super-resolution imaging. By properly engineering the point spread function (PSF), such as utilizing Laguerre-Gaussian (LG) modes and their superposition, the ultimate limits of 3D localization precision can be enhanced. However, achieving these limits is challenging, as it often involves complicated detection strategies and practical limitations. In this work, we rigorously derive the ultimate 3D localization limits of LG modes and their superposition, specifically rotation modes, in the multi-parameter estimation framework. Our findings reveal that a significant portion of the information required for achieving 3D super-localization of LG modes can be obtained through feasible intensity detection. Moreover, the 3D ultimate precision can be achieved when the azimuthal index $l$ is zero. To provide a proof-of-principle demonstration, we develop an iterative maximum likelihood estimation (MLE) algorithm that converges to the 3D position of a point source, considering the pixelation and detector noise. The experimental implementation exhibits an improvement of up to two-fold in lateral localization precision and up to twenty-fold in axial localization precision when using LG modes compared to Gaussian mode. We also showcase the superior axial localization capability of the rotation mode within the near-focus region, effectively overcoming the limitations encountered by single LG modes. Notably, in the presence of realistic aberration, the algorithm robustly achieves the Cram\'{e}r-Rao lower bound. Our findings provide valuable insights for evaluating and optimizing the achievable 3D localization precision, which will facilitate the advancements in super-resolution microscopy.

\textbf{Keywords}: 3D localization, multi-parameter estimation, maximum likelihood estimate, Laguerre-Gaussian modes, quantum Fisher information
\end{abstract}

\maketitle


\section{\label{sec:level1}Introduction}
Precise three-dimensional (3D) localization is essential for a variety of advanced microscopy techniques, including defect-based sensing \cite{chen2015subdiffraction,jaskula2017superresolution}, multiplane detection \cite{dalgarno2010multiplane,abrahamsson2013fast}, single-particle tracking \cite{manzo2015review, Three2017, shen2017single}. Especially, as the position of individual fluorophores can be determined with a precision smaller than the size of the point spread function (PSF), a super-resolution image can be assembled from the estimated positions of the sufficient fluorescent labels \cite{Breaking1994,imaging2006,rust2006stochastic, Three2008}, to avoid the Abbe-Rayleigh criterion \cite{XXXI1879, Principles1999}. The effective achievable resolution is closely related to the precision of individual fluorophore localization. The 3D localization is intimately connected to multi-parameter estimation. Therefore, the advantages of these techniques are better understood in the multi-parameter framework.

 Building upon the pioneering work of Tsang and coworkers in quantifying far-field two-point super-resolution \cite{tsang2016quantum, tsang2017subdiffraction, tsang2018subdiffraction}, the ultimate precision limits of single point source's localization have been extensively investigated \cite{Quantum2015, Fundamental2018, yu2018quantum}.  The basic idea is to exploit the quantum Fisher information (QFI) and associated quantum Cram\'{e}r-Rao bound (QCRB) \cite{Quantum1969, Statistical1994}. Its theory proves to be effective in establishing under which conditions systems may be preferable to estimate parameters. Common imaging strategies rely on complicated experimental setups to achieve the quantum-mechanical bound, such as interferometer arrangement \cite{yang2016far, tang2016fault}, mode sorter \cite{dutton2019attaining, zhou2019quantum}, and spatial-mode demultiplexing (SPADE). The multi-parameter cases often involve trade-off relations among the uncertainties on the parameter, since the total information has now to be apportioned \cite{liu2020quantum, albarelli2020perspective}. This makes the multi-parameter optimization problem more involved and more intriguing.

Given the close relationship between QFI and the characteristics of the light field, precision can be significantly enhanced by modifying the system's response, such as through PSF engineering \cite{Nanometric2007, High2008, Three2009}. Recent studies have demonstrated the immense potential of LG modes, as opposed to conventional Gaussian mode, for improving 3D localization precision \cite{Quantum2021}. The superposition of LG modes, specifically the rotation mode characterized by their intensity profiles that rotate on propagation, further enhances the ultimate precision of axial localization. Moreover, the ultimate axial precision of LG modes can be achieved with intensity detection \cite{Intensity2019, Axial2021}. The simplicity and feasibility of intensity detection make it extremely valuable for microscopy applications, circumventing potential systematic errors and losses inherent in complex strategies delineated previously \cite{linowski2023application}. However, practical intensity detection introduces pixelated readouts and inherent detection noise, which can compromise the theoretical advantages offered by this simple optimal strategy. Therefore, rigorous mathematical analysis and robust localization algorithms are necessary.

In this work, we rigorously derive the ultimate 3D localization limits of LG modes and rotation modes in the multi-parameter estimation framework. We investigate the accessibility of these limits under ideal intensity detection conditions. Furthermore, we consider the practical limitations imposed by finite pixel size and various detector noise sources. Taking these factors into account, we develop a robust maximum likelihood estimation (MLE) algorithm that iteratively determines the 3D position of a point source.  Our algorithm robustly achieves the CRB under low signal-to-noise ratio (SNR) and aberrational conditions. Moreover, it is not limited to symmetric PSFs, but can be extended to accommodate anisotropic PSFs, as well as diverse noise statistics. In this manner, we present comprehensive theoretical and experimental evidence aiming at exhibiting remarkable super-localization capabilities facilitated by LG and rotation modes.

\section{\label{sec:level2}Theoretical framework}

\begin{figure}[bt]
\includegraphics[width=0.48\textwidth]{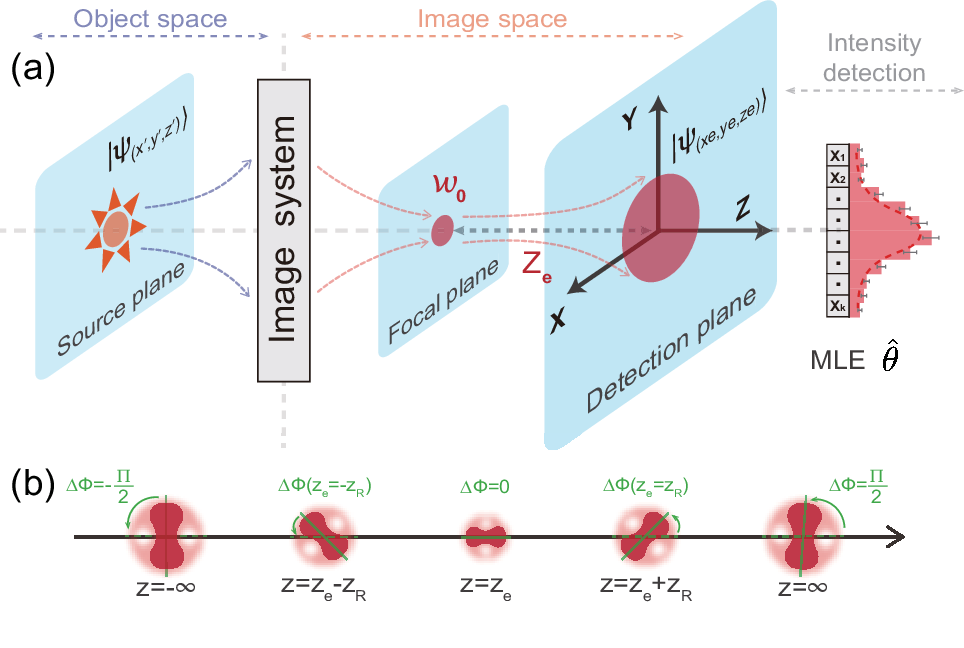}
\caption{\label{fig: schematic} (a) Schematic illustration of the 3D localization of a point source with an optical microscope-based imaging setup. (b) Schematic illustration of the lateral intensity profiles variation with the rotation mode at different axial positions.}
\end{figure}

Localization can be regarded as a multi-parameter estimation problem, aiming to determine the 3D coordinates of a point source in the image space, as depicted in Fig.~\ref{fig: schematic}(a). Assuming an initial state denoted by $\vert\Psi_{(0)}\rangle$ in the image space, the 3D displacement can be described by a unitary operation:
\begin{equation}
\label{pure state}
    \vert\tilde{\Psi}_{(x_e,y_e,z_e)}\rangle=\exp (-i \hat{G}_z z_e-i \hat{p}_{x} x_e-i \hat{p}_{y} y_e)\vert\Psi_{(0)}\rangle.
\end{equation}
Here, the operators $\hat{p}_{x}=-i\partial_x$ and $\hat{p}_{y}=-i\partial_y$ represent the lateral displacement as momentum operators, and the axial displace operator is  denoted as $\hat{G}_z=-i\partial_z=\frac{1}{2 k} \nabla_{T}^{2}+k$, where $k$ is the wavenumber and $\nabla_{T}^{2}=\partial_{x x}+\partial_{y y}$. These operators commute with each other, enabling simultaneous measurement of these unknown parameters \cite{New2002, Compatibility2016}. Through the aforementioned approach, the state in the image space, represented by $\rho_\theta=\vert\tilde{\Psi}\rangle\langle\tilde{\Psi}\vert$, is parameterized with point source 3D coordinates $\theta=(x_e,y_e,z_e)$. The image is a magnified replica of the state $\vert\Psi_{(x',y',z')}\rangle$ in the object space~\cite{Fundamentals2019}. The estimation of 3D coordinates of the object requires a parametric transformation related to the magnification factor of the system.
 
To proceed further, we consider a shifted normalized $LG$ mode, with a transverse field  in the detection plane, given by \cite{Introduction2005} 
\begin{equation}
\label{wave function}
\begin{aligned}
&LG_{lp}(r,\phi,z)=\langle x,y,z\vert\tilde{\Psi}_{(x_e,y_e,z_e)}\rangle=\sqrt{\frac{2 p !}{\pi(p+\vert l\vert) !}}\\
  &\times \frac{1}{w(z)} \left[\frac{\sqrt{2r^2}}{w(z)}\right]^{\vert l \vert}L_{p}^{\vert l \vert} \left[\frac{2 r^2}{w(z)^{2}}\right] \exp \left[-\frac{r^2}{w\left(z\right)^{2}}\right] \\
  &\times \exp\left[-i\left(k z-\frac{kr^2}{2 R(z)}+\vert l\vert\phi+\zeta_{lp}(z)\right)\right],
\end{aligned}
\end{equation}
where $r^2=(x-x_e)^2+(y-y_e)^2$, $\phi=(x-x_e)/(y-y_e)$, $L_p^{\vert l \vert} [\cdots]$ is the generalized Laguerre polynomial defined by azimuthal index $l\in\mathbb{Z}$ and radial index $p \in\mathbb{Z}^{+}$. The transverse field distribution is determined by the beam waist radius $w_0$ and the Rayleigh range $z_R$ through $R(z)=(z-z_e)\left[1+( \frac{z_{\mathrm{R}}}{z-z_e})^{2}\right]$, $w^{2}(z)=w_{0}^{2}\left[1+(\frac{z-z_e}{z_{\mathrm{R}}})^{2}\right]$, $\zeta_{lp}(z)=(2 p+\vert l\vert+1) \arctan (\frac{z-z_e}{z_{\mathrm{R}}})$ and $z_R=\frac{\pi w_0^2}{\lambda}$. To streamline the derivation, we redefine the coordinate system with the  detection plane serving as the origin, denoted as $z=0$ and $z_e$ represents the distance between the detection plane and the focal plane.

The 3D localization precision is quantified by the covariance matrix $Cov(\boldsymbol{\Pi}, \tilde{\theta})$ of locally unbiased estimators, which is lower bounded by Cram\'{e}r-Rao bound (CRB) and QCRB:
\begin{equation}
    Cov(\boldsymbol{\Pi}, \tilde{\theta}) \ge   \frac{1}{N} \mathcal{F}\left(\rho_{\theta}, \boldsymbol{\Pi}\right)^{-1}\ge   \frac{1}{N} \mathcal{Q}\left(\rho_{\theta}\right)^{-1},
\end{equation}
where  $N$ is the number of system copies related to the effective photon counts in each frame. The associated classical Fisher information matrix (CFIm), denoted as $\mathcal{F}\left(\rho_{\theta}, \boldsymbol{\Pi}\right)$ is defined by
\begin{equation}
\label{FI}
  \mathcal{F}_{ij}\left(\rho_{\theta}, \boldsymbol{\Pi}\right)=\sum_{X_{k}} \frac{1}{P\left(X_{k} \vert\theta\right)}\frac{\partial  P\left(X_{k} \vert\theta \right)}{\partial \theta_i} \frac{\partial  P\left(X_{k} \vert\theta \right)}{\partial \theta_j},  
\end{equation}
where $P\left(X_{k} \vert\theta\right)$ is the conditional probability density of observing an outcome $X_k$ depending on the underlying 3D position $\theta$ of the source, and a specific measurement $\boldsymbol{\Pi}$. The associated  QFI matrix (QFIm), denoted as $\mathcal{Q}(\rho_\theta)$, gives the maximum of the CFIm. In the case of pure states, as is the situation we consider, the QFIm is four times the covariance matrix of the generators, which consists solely of diagonal entries.

The axial QFI for arbitrary LG modes has been recently worked out in Ref.~\cite{Axial2021}, we extend the results into a 3D scenario, which can be expressed as:
\begin{equation}
\label{QFI}
   \mathcal{Q}_{x,y}=\frac{4(2p+\vert l\vert+1)}{w_0^2},\mathcal{Q}_{z}=\frac{2p(p+\vert l \vert)+2p+\vert l \vert +1}{z_R^2}.
\end{equation}
The lateral QFI exhibits a linear dependence on $p$, while the axial QFI shows a quadratic dependence on $p$, highlighting the significant role of the radial index in effectively enhancing 3D localization precision.
These findings are consistent with the numerical results presented in Ref.~\cite{Quantum2021} for low-order LG modes. Notably, the $LG_{00}$ mode is the Gaussian mode, serving as a classical benchmark for comparative analysis. 

While LG modes exhibit trivial divergence during propagation, more intricate intensity transformations can be achieved by superposing different LG modes. In particular, when employing a set of $M$ constituent LG modes that satisfy the relation $[(2p+l)_{j+1}-(2p+l)_j]/[l_{j+1}-l_{j}]\equiv const \equiv V$ for $j=1,2,..., M-1$, the resulting intensity pattern exhibits anisotropy (with non-circular symmetry) and undergoes rotation during the propagation \cite{Wave1996, Propagation2000}. As illustrated in Fig.\ref{fig: schematic}(b), the overall rotation angle from the waist to the far field is given by $\Delta\phi(z_e=\infty)=V\pi/2$, and $\Delta \phi(z_e=-\infty)=-V\pi/2$. Notably, half of $\Delta \phi$ is obtained at the Rayleigh range. These PSFs offer a wider range of applications in super-resolution imaging due to their superior localization precision compared to circular symmetric PSFs \cite{Three2008, Optimal2014, Depth2006}. To provide a clear physical intuition, we consider the simple example of the superposition of two LG modes (M=2) with $l\neq l^{\prime}$ and $p=p^{\prime}=0$. The 3D QFI for rotation modes can be determined as follows:
\begin{equation}
\begin{aligned}
    \mathcal{Q}_{x,y}&=\frac{2(\vert l\vert+\vert l^{\prime}\vert+2)}{w_0^2},\\
    \mathcal{Q}_{z}&=\frac{[4+2(|l|+|l'|)+(|l|-|l'|)^2]}{z_R^2}.\\ 
\end{aligned}
\end{equation}
In contrast to the quadratic precision improvement in the axial localization, the lateral QFI is obtained by summing up the contributions from individual modes.

\begin{figure}[!htbp]
\includegraphics[width=0.5\textwidth]{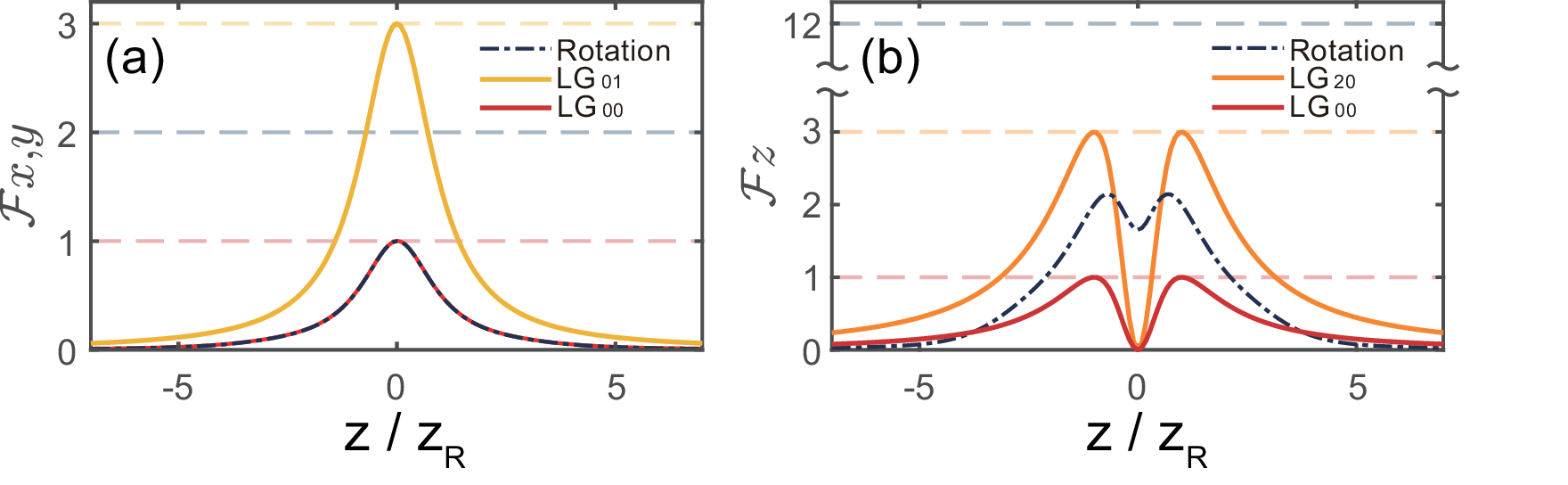}
\caption{\label{fig: ratio} The (a) lateral CFI and (b) axial CFI for ideal intensity detection as a function of the distance between the detection plane and the focal plane. The average lateral CFI of the rotation mode is equivalent to that of $LG_{00}$. The horizontal dashed lines indicate the QFI of each mode. Each curve is normalized to the QFI of $LG_{00}$.}
\end{figure}

Typically, the QFI is distributed between the phase and intensity variations of the measured beam. Remarkably, by discarding the phase information, the full axial QFI can still be extracted \cite{Axial2021}. This result prompts us to investigate the efficacy of intensity detection in achieving ultimate 3D localization precision. When ideal detection conditions are assumed, i.e. a detector of infinite area without pixelation and no additional noise sources except for shot noise, the intensity detection projects the quantum state onto the eigenstates of spatial coordinates, represented as $\boldsymbol{\Pi}_{x, y}=\vert x, y\rangle\langle x, y\vert$. In consequence, the probability density is $P(X_k\vert \theta )=\operatorname{Tr} (\rho_\theta \boldsymbol{\Pi}_{x, y})=\vert \tilde{\Psi}_{(x_e,y_e,z_e)}\vert ^{2}$, which corresponds to the (normalized) beam intensity $\vert LG_{lp}\vert^2$. The summation in Eq.~\ref{FI} transforms into a two-dimensional integral over the spatial domain. For LG modes, after a lengthy calculation, the ideal CFI can be expressed analytically as follows:
\begin{equation}
    \mathcal{F}_{x,y}(z_e)=\frac{4(2p+1)}{{w(z_e)}^{2}},\mathcal{F}_z(z_e)=\frac{4[2p(p+\vert l \vert)+2p+\vert l \vert +1]}{{R(z_e)}^{2}}.
\end{equation}
These results are plotted in Fig.~\ref{fig: ratio}. Two detection planes can be found, where full axial QFI and a portion of lateral QFI can be extracted. In the case of certain LG modes with $\vert l \vert=0$, the lateral CFI can reach the QFI at the beam waist, $\mathcal{F}_{x,y}(0)=\mathcal{Q}_{x,y}$, while the axial CFI can reach it at the Rayleigh range, $\mathcal{F}_{z}(\pm z_R)=\mathcal{Q}_{z}$. Intuitively, the precision of axial localization depends on the beam divergence, which determines the rate of beam width variations during propagation. The precision of lateral localization benefits from sharpness of the wave packet. By increasing the radial index $p$, the sharpness and the beam divergence are enhanced \cite{wan2022divergence}, leading to an improvement in the precision of 3D localization. However, as the azimuthal index $l$ is increased, a central dark spot emerges and expands in size. Although this leads to increased divergence, it fails to enhance the sharpness of the beam, thereby hindering improvements in the precision of lateral localization. In the case of rotation modes, we consider the superposition of $LG_{00}$ and $LG_{20}$ modes as a representative example. Numerical analysis suggests that only a small fraction of the 3D QFI can be extracted with intensity detection. As this specific category lacks radial information, the average lateral CFI is equivalent to that of the $LG_{00}$, see Fig.~\ref{fig: ratio}(a). However, the non-stationary rotation behavior significantly enhances axial CFI in the near-focus axial region compared to single LG mode, as shown in Fig.~\ref{fig: ratio}(b). These results also underscore the significance of developing quantum-inspired strategies, such as SPADE or mode sorting techniques, to reveal all information about the parameter. Comprehensive derivations of the QFIm and ideal CFIm are included in the Appendix \ref{calculate QFI CFI}.

We then incorporate the deteriorating effects in the detection process and present our MLE algorithm. Previous studies have demonstrated that the localization algorithms based on MLE can asymptotically approach the CRB for a few specific scenarios \cite{Localization2004, Stochastic2006, Fast2010}, outperforming nonlinear least squares (NLLS) algorithm \cite{Quantitative2009}. However, discrepancies between the variances of MLE and the precision predicted by the CRB have been observed in the presence of model mismatches and misspecifications \cite{Fast2010, zhou2019quantum}, such as inaccurate noise statistics, PSF mismatch, optical aberrations, and low SNR. These limitations stem from the fact that existing localization MLE algorithms make simplified statistical assumptions for specific scenarios, which inspires us to improve the robustness and generalisability of MLE algorithms by adopting a more refined statistical model. In the case of pixelated intensity detection, the measurements can be described as $\boldsymbol{\Pi}_{A_k} = \int_{A_k} \vert x, y\rangle\langle x, y\vert dxdy $. The readout counts in the $k$th pixel encompass the integrated photon counts $N_k$ and contributions from detector noise, which can be expressed as:
\begin{equation}
\label{readout}
X_k = N\int_{A_k}\vert\tilde{\Psi}_{(x_e,y_e,z_e)}\vert ^{2}dxdy + N_{b} + N_{c}.
\end{equation}
The photon-electric conversion process in the CCD camera distorts the effective photon counts, resulting in two types of detection noise. The term $N_b$ represents signal-independent noise, including background fluorescence, dark current, and readout noise. On the other hand, the variance of $N_c$ positively correlates with the signal and arises in the electron amplification process \cite{janesick1987scientific}. The SNR is determined by the ratio of the average effective photon counts to the noise present at each pixel. These statistical assumptions have been successfully applied in weak measurement scenarios with limited SNR and detector dynamic range 
 \cite{Approaching2020, Improving2021}.

Based on the statistical model mentioned above, we describe the MLE algorithm to estimate the parameters with the likelihood function:
\begin{equation}
\label{likehood}
\mathrm{ln} \textbf{L}(\vec{X}\vert\theta) = \sum_k \mathrm{ln} P(X_k\vert\theta).
\end{equation}
The estimated results, denoted as $\hat{\theta}$, maximize the likelihood function. We employ the scoring method to iteratively update the parameter estimates using the inverse of the CFIm and the derivative of the likelihood function:
\begin{equation}
\hat{\theta}_{n+1} = \hat{\theta}_n + \mathcal{F}(\rho_\theta, \boldsymbol{\Pi}_{A_k})^{-1} \frac{\partial \mathrm{ln} \textbf{L}(\vec{X}\vert\theta)}{\partial\theta}\bigg\vert_{\theta=\hat{\theta}_n}.
\end{equation}
The iterative update scheme is similar to previous approaches in Refs.~\cite{aguet2005maximum, Fast2010}, but improves iteration stability and reduces computational complexity \cite{Fundamentals1993}.

For a system with unknown $w_0$, the algorithm simultaneously estimates three unknown parameters: $\theta=(x_e,y_e,w(z_e))$.  We assume the nominal axial distance is known, and the estimated beam width is used to determine the system's $w_0$ and $z_R$. The lateral variance $var(\hat{x}_e,\hat{y}_e)$ can be directly calculated, while axial $var(\hat{z}_e)$ can be derived using error propagation $var(\hat{z}_e)=var(\hat{w}(z_e))/[\partial_{z_e} \hat{w}(z_e)]^2$. If the system is fully pre-calibrated with a known $w_0$, the algorithm can estimate $z_e$ instead of $w(z_e)$ with slight modification to handle anisotropic PSF, enabling direct 3D position estimation, as demonstrated in the following rotation mode experiment.

\section{Experiment}
\subsection{Experiment setups}
\begin{figure*}[t]
\includegraphics[width=0.87\textwidth]{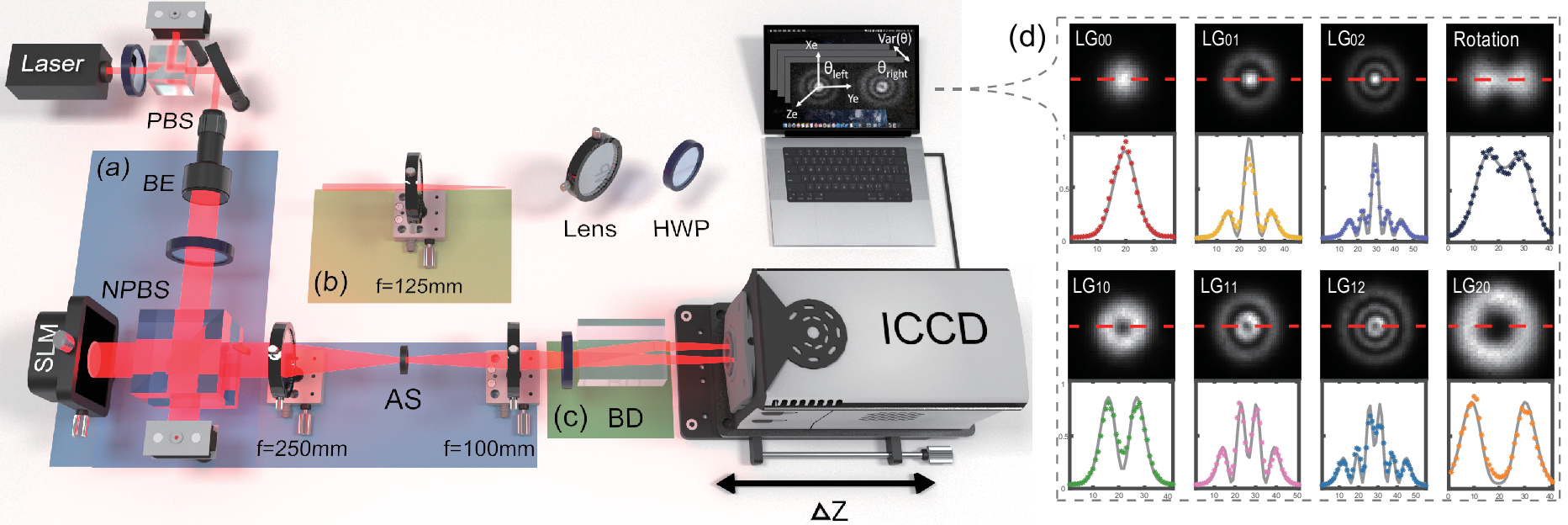}
\caption{\label{fig: exp} Experimental setups for 3D localization of a point source. The notations used are as follows: (N)PBS, (non) polarized beam splitter; BE, beam expander; SLM, spatial light modulator; AS, aperture stop; HWP, half-wave plate; BD, beam displacer; ICCD, intensified charge-coupled devices. A detailed description of the functions performed by modules (a)-(c) can be found in the text. (d) Measured intensity patterns  with cross sections (asterisk) and corresponding fitted profiles (solid curves). }
\end{figure*}

\begin{figure}[b]
\includegraphics[width=0.48\textwidth]{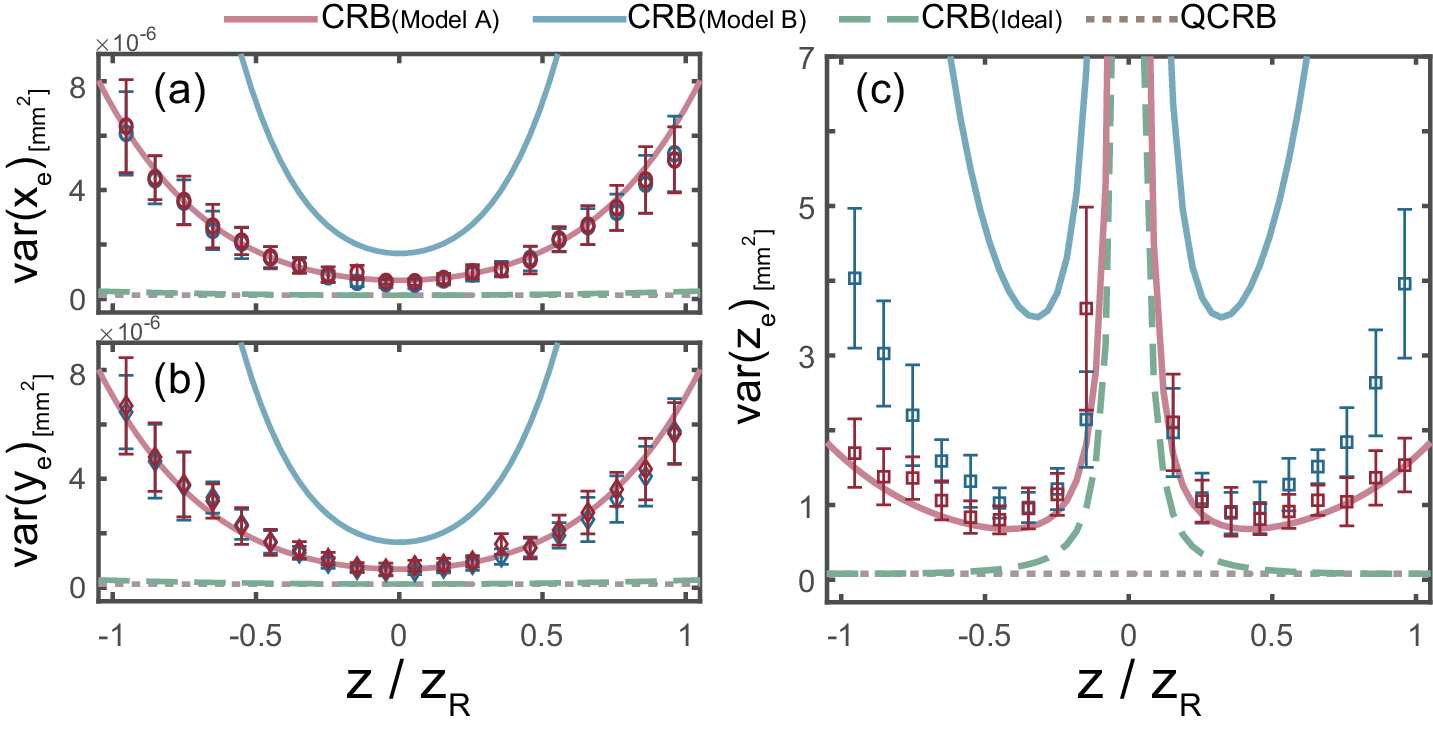}
\caption{\label{fig: Gau} Comparison between the 3D localization precision of imperfect Gaussian mode for (a) $x_e$ (b) $y_e$ and (c) $z_e$ position using different MLE algorithms. The theoretical results (lines) are determined by the CRB and the experimental results (points) are obtained using MLE.}
\end{figure}

\begin{figure*}[tb]
\includegraphics[width=1\textwidth]{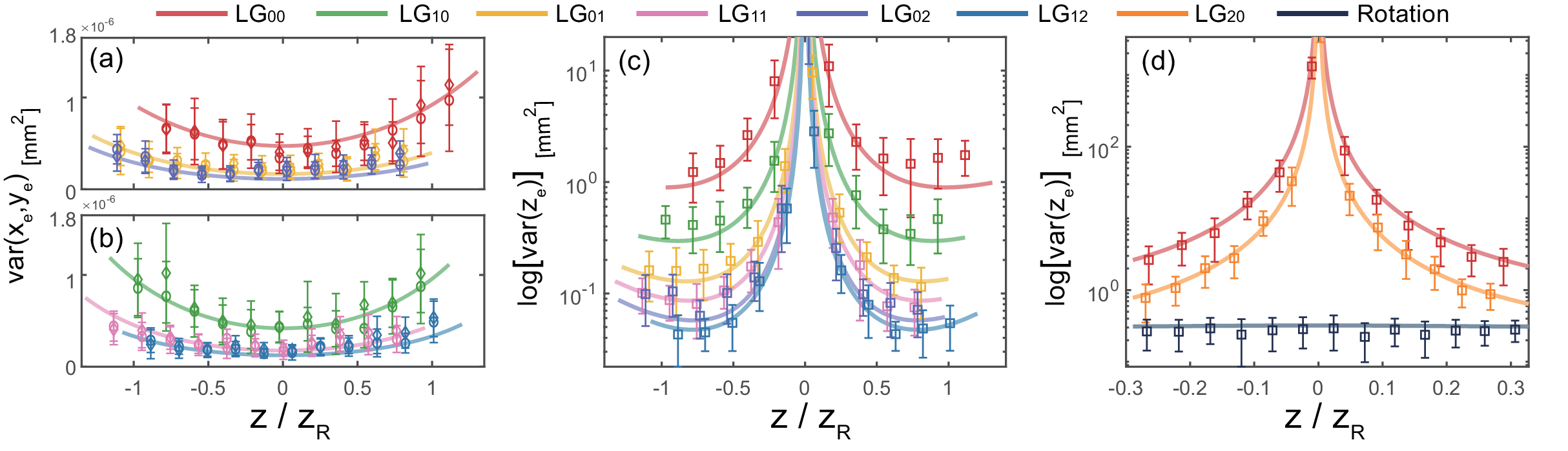}
\caption{\label{fig: LG} Experimental results of LG modes and the rotation mode with the lateral localization precision of (a) $LG_{l=0,p=\{0,1,2\}}$ modes and (b) $LG_{l=1,p=\{0,1,2\}}$ modes as well as the axial localization precision of (c) $LG_{l=\{0,1\},p=\{0,1,2\}}$ modes and (d) the rotation mode. The theoretical results (lines) are determined by the CRB and the experimental results (points) are obtained using MLE.}
\end{figure*}

To validate our theoretical framework, we conducted experimental 3D localization using imperfect Gaussian mode, LG modes, and the rotation mode. The experimental setup is schematized in Fig.~\ref{fig: exp}. In these experiments, we use the focus beam waist as a simplified point source realization.

We first assess the performance of the MLE algorithm in the presence of model mismatches and misspecifications by 3D localizing an imperfect Gaussian mode. A He-Ne laser at wavelength $\lambda=633 \text{nm}$ serves as the Gaussian source. A single lens in Module (b) is utilized to form the image, which leads to a beam waist $w_0=77.48\mu\text{m}$ and corresponding Rayleigh range $z_R=29.8\text{mm}$. The spatially unfiltered laser and the spherical aberration of the image system lead to the imperfection of Gaussian mode \cite{Spherical2018}. A sequence of images is captured at different axial positions using a scientific ICCD camera (Andor, iStar CCD 05577H) with pixel size $13\mu \text{m} \times 13\mu \text{m}$. The axial positions range over a span of $60 \text{mm}$ with an interval of $3 \text{mm}$. At each position, we acquire $600$ intensity images. The pre-calibration noise are characterized by $N_b\sim \mathcal{N}\left(515.6, 7.1^{2}\right)$ and $N_c\sim \mathcal{N}\left(0, \sigma_{c}^{2}\right)$. Here, $\ln \left(\sigma_{c}^{2}\right)=1.4 \ln \left(N_k\right)-0.7$ and $N_k$ represents the effective photon counts per pixel. It is worth noting that while Gaussian noise aligns with our CCD response calibration, other statistical distributions can also be accommodated in the algorithm. The effective photon counts per image are approximately $N=1\times10^4$, obtained by subtracting the detector noise from the total readout. These conditions reflect the typically low SNR encountered in real microscopy scenarios.

We then compare the 3D localization precision of LG modes and the rotation mode, demonstrating their super-localization capabilities. The desired PSFs are generated using a double-fourier transform optical setup, as depicted in Module (a). Computer-generated holograms (CGH) are imprinted onto the SLM (Meadowlark Optics, P1920-400-800-HDMI), with the desired first-order diffraction selected by a 4f system \cite{ Comparison2016, Pixelated2007, Mode2009}. The waist radius of CGHs is uniformly set to $500\mu \text{m}$. The second lens of the 4-f system is slightly displaced from the focal plane to ensure proper beam focusing. The modulation efficiency of the SLM is adjusted by an HWP. To mitigate the adverse impacts of beam jitter and turbulence, an additional HWP and a BD in Module (c) are employed to create two copies of the output beam, namely the $left$ and $right$ beams. These negative effects have less impact considering the previous single-lens imaging systems. By subtracting the estimation results obtained from two replicated beams, we obtain the final variance $var(\hat{\theta})=var(\hat{\theta}_{left}-\hat{\theta}_{right})/2$. To ensure a fair comparison, the different modes are normalized to have the same effective photon counts. Each stack is comprised of $200$ images. An overview of the measured (normalized) intensity patterns is summarized in Fig.~\ref{fig: exp}(d). The model mismatch is characterized by the deviation from the ideal lateral intensity distribution and beam divergence, as discussed in detail in the Appendix \ref{model mismatch}. The results indicate that the imperfect Gaussian mode exhibits a severe model mismatch compared to modes generated through the 4f system.
 
The 3D localization precision is quantified by the covariance matrix. As the matrix contains only diagonal entries, the variance of each estimator is sufficient to measure the precision. To obtain error bars for the variances, we divide the raw data into 20 groups assuming repeated experiments. 

\subsection{Experiment results}
The experimental results related to imperfect Gaussian mode are summarized in Fig.~\ref{fig: Gau}. Our algorithm is referred to as Model A. As a comparison, we employ another widely used localization MLE algorithm, referred to as Model B, as described in Ref.~\cite{Fast2010}. The key distinction between the two models lies in Model B's consideration of a Poisson noise source $N_b$ while neglecting the contribution of $N_c$. Additionally, Model B treats $N$ and $N_b$ as unknown parameters, and offers direct estimation of $\theta=(x_e,y_e,w(z_e),N,N_b)$. As the sub-CFIm of $\theta=(w(z_e), N, N_b)$ contains non-diagonal entries, $N$ and $N_b$ can often be thought of nuisance parameters \cite{suzuki2020quantum}. 

As illustrated in Fig.~\ref{fig: Gau}, the ultimate localization precision is given by the QCRB, while CRB of ideal intensity detection reaches QCRB at the focus and the Rayleigh range. The presence of noise diminishes the precision achievable with the ideal CRB. To address this issue, the practical CRB can be obtained using the statistical models, as discussed in detail in the Appendix \ref{calculate NCRB}. With increasing the distance between the detection plane and the focal plane, the SNR decreases, widening the gap between the practical CRB and the ideal one. By employing a more refined statistical model, Model A accurately predicts the CRB. Conversely, Model B provides a significantly non-tight CRB due to noise misspecification, specifically the Poisson assumption overestimating the variance of the noise. Both algorithms obtain similar precision in the lateral direction, see Fig.~\ref{fig: Gau}(a)-(b). However, as the distance increased, substantial discrepancies in the variance of axial localization become apparent, as shown in Fig.~\ref{fig: Gau}(c). We infer that these discrepancies may be attributed to nuisance parameters and model mismatch, which often result in reduced algorithm precision \cite{Quantitative2009, rosati2023fundamental}. The inaccurate estimation of $w(z_e)$ occurs due to the biased estimations of $N$ and $N_b$ when there exist PSF mismatches caused by aberrations, as discussed in the Appendix \ref{model mismatch}. The limitation of Model B in achieving the axial CRB has also been observed in previous works \cite{Fast2010, zhou2019quantum}. These results align with our intuition that a more refined model, taking into account more characteristics of the experimental apparatus, enhances the algorithm's precision and robustness against model misspecification and mismatches. 

The experimental results of 3D localizing a set of LG modes are presented in Fig.~\ref{fig: LG}(a)-(c). The images are captured at $10\mu\text{m}$ intervals within a $100\mu\text{m}$ range, while the Rayleigh range is approximately $52.71\mu\text{m}$. By ensuring modulation accuracy and efficiency, continual improvement in 3D localization precision can be achieved using higher-order modes. Specifically, for the highest-order mode $LG_{12}$ generated in our experiment, the variance of the lateral localization is $0.86\times 10^{-3} \text{mm}^2$ and the variance of the axial localization is $0.25\times 10^{3} \text{mm}^2$. In comparison, the $LG_{00}$ mode, which serves as the classical benchmark, achieves a lateral localization precision of $2.03\times 10^{-3} \text{mm}^2$ and an axial localization precision of $7.28\times 10^{3} \text{mm}^2$ under the same detection conditions. The results exhibit an enhancement of up to two-fold in lateral localization precision and up to twenty-fold in axial localization precision when employing LG modes.  However, the deleterious effects of pixelation and noise will be exacerbated in higher-order modes. It is imperative to assess these potential limitations in order to achieve the anticipated precision improvement. Additionally, we present the results of the axial localization experiment for the rotation mode in Fig.~\ref{fig: LG}(d), along with the constituent modes $LG_{00}$ and $LG_{20}$.  Instead of changing the axial distance of the detector plane, a series of holograms are utilized to simulate the propagation of the light field. By performing a pre-calibration of the beam waist, our algorithm enables direct estimation of 3D coordinates. In the vicinity of the focal plane, the CFI of LG modes approaches zero, with their lateral intensity distributions seldom changing during propagation. Moreover, the likelihood function of LG modes exhibits the same values in two axial positions, leading to an ambiguous axial position estimate of $\pm z$. In contrast, rotation modes exhibit a unique and easily detectable rotation angle $\Delta \phi(z_e)$ as they propagate \cite{Wave1996, Axial2021}. This characteristic eliminates the ambiguity of $\pm z$ and greatly improves the precision of axial localization. The experimental results highlight the exceptional precision in the axial localization achieved with the rotation mode in the near-focus region, surpassing that of the LG modes.

\section{Conclusion}
In summary, our research provides both theoretical and experimental evidence showcasing the exceptional potential of LG and rotation modes for 3D super-localization. To address practical challenges, we develop an iterative MLE algorithm that effectively estimates the 3D positions of point sources with the best possible precision  determined by the CRB. By incorporating a refined noise statistic model, our algorithm improves the robustness and generalizability of the localization process, offering significant advantages in scenarios with low SNR and aberrations.

While higher-order or intricate superposition modes demonstrate theoretical advantages in ideal intensity detection, practical experimental imperfections pose challenges in realizing these benefits with PSF engineering and intensity-based strategies.  Therefore, when exploring and optimizing the final effective resolution, the practical CRB provided by our algorithm can serve as a reliable benchmark for evaluating precision improvements. Our work builds a bridge between the quantum estimation framework and practical microscopy algorithm,  fostering promising advancements in 3D super-resolution microscopy.

\subsection*{Acknowledgments}
We thank Mingnan Zhao for the helpful discussions in deriving the classical Fisher information matrix. 

\subsection*{Funding}
This work is supported by the National Natural Science Foundation of China (Grants No. 61975077), the National Key Research and Development Program of China (Grants No. 2019YFA0308704), Civil Aerospace Technology Research Project (Grants No. D050105).

\subsection*{Availability of data and materials}
All the data and materials relevant to this study are available from the corresponding author upon reasonable request.

\section*{Declarations}
\subsection*{Ethics approval and consent to participate}
Not applicable.
\subsection*{Consent for publication}
Not applicable.
\subsection*{Competing interests}
All authors declare that there are no competing interests.
\subsection*{Author contributions}
CH did the experiments with the help of LX, BW, YZ1 and ZL. YZ2 and LZ supervise the project. All authors read and approved the final manuscript.

\appendix
\section{Derivation of Quantum and Classical Fisher Information Matrix }
\label{calculate QFI CFI}
Paraxial wave optics has several analogies with fundamental quantum mechanics, the precursor idea for deriving the QFIm is to link the paraxial wave equation with the Schr\"odinger equation for the wave function of a two-dimensional space. LG modes correspond to the stationary harmonic oscillators eigenstates that can be obtained by applying the circular ladder operators~\cite{Analogies2017, Paraxial1993}:
\begin{equation}
    \hat{a}_{\pm} =\frac{1}{\sqrt{2}}\left(\hat{a}_{\xi} \mp i \hat{a}_{\eta}\right).
\end{equation}
The dimensionless operators $\hat{a}_{\xi}$ and $\hat{a}_{\eta}$ correspond to the independent amplitudes of the oscillator and follow the bosonic commutation rules $[\hat{a}_{s},\hat{a}_{s'}^\dagger]=\delta_{ss'}$ ($s,s'\in\xi,\eta$). The operators $a_\pm$ also obey similar relations. These operators act on the vacuum state to generate the eigenstates of the harmonic oscillator, denoted as $\vert \textit{n}_+,\textit{n}_-\rangle$. These eigenstates correspond to LG modes, which are characterized by the azimuthal index $l$ and radial index $p$:
\begin{equation}
     l=n_{+}-n_{-}, \quad p=\min \left(n_{+}, n_{-}\right).
\end{equation}

The definition of lateral displacement generator $\hat{p}$ and axial displacement generator $\nabla_T^2$ are:
\begin{equation}
   \begin{aligned}
     \hat{p}_{\xi}&=\frac{1}{\sqrt{2}\textit{i}}(\hat{a}_\xi-\hat{a}_\xi^\dagger)=\frac{\textit{i}}{2}(\hat{a}_+^\dagger+\hat{a}_-^\dagger-\hat{a}_+-\hat{a}_-),\\ 
     \hat{p}_{\eta}&=\frac{1}{\sqrt{2}\textit{i}}(\hat{a}_\eta-\hat{a}_\eta^\dagger)=\frac{1}{2}(\hat{a}_+^\dagger-\hat{a}_-^\dagger+\hat{a}_+-\hat{a}_-),\\
      {\nabla}_{T}^{2}&=p_{\xi}^{2}+p_{\eta}^{2}=\left(p_{\xi}+\textit{i} p_{\eta}\right)\left(p_{\xi}-\textit{i} p_{\eta}\right)=\left(a_{+}-a_{-}^{\dagger}\right)\left(a_{+}^{\dagger}-a_{-}\right).
   \end{aligned}    
\end{equation}

Considering in a proper unit, $(x,y,\nabla_{T}^2){\mapsto}(\sqrt{2}\xi/w_0,\sqrt{2}\eta/w_0,2\widetilde{\nabla}_T^2/w_0^2)$, we obtain $(\hat{p}_x,\hat{p}_y,\hat{G}_z){\mapsto}(2\hat{p}_\xi/w_0^2,2\hat{p}_\eta/w_0^2,\widetilde{\nabla}_T^2/4z_R^2)$. In the case of pure states, the QFIm can be expressed as four times the covariance matrix of the generators computed in eigenstates $\vert \textit{n}_+,\textit{n}_-\rangle$, computed in the eigenstates $\vert \textit{n}+,\textit{n}-\rangle$. The entries of the QFIm for pure LG modes are given by
\begin{equation}
\label{qfix}
\begin{aligned}
  Q_{xx}&=4 \left(\langle \textit{n}_+,\textit{n}_- \lvert \hat{p}_x^2\vert\textit{n}_+,\textit{n}_- \rangle - \langle \textit{n}_+,\textit{n}_- \lvert \hat{p}_x\vert\textit{n}_+,\textit{n}_- \rangle^2 \right)\\
&=\frac{4}{\textit{w}_0^2}(n_++n_-+1)\\
&=\frac{4}{\textit{w}_0^2} \left( 2p+\vert l\vert+1 \right)\\
&=Q_{yy},\\
Q_{zz}&=4 \big( \langle \textit{n}_+,\textit{n}_- \lvert  \hat{G}^2\vert\textit{n}_+,\textit{n}_- \rangle-\langle \textit{n}_+,\textit{n}_- \lvert  \hat{G}\vert\textit{n}_+,\textit{n}_- \rangle^2 \big) \\
&=\frac{4}{z_R^2}\left(2 n_{+} n_{-}+n_{+}+n_{-}+1\right)\\
&=\frac{4}{z_R^2}[2 p(p+\vert l\vert)+2 p+\vert l\vert +1].\\
Q_{xz}&=4 \big(\langle \hat{p}_x  \hat{G} \rangle - \langle \hat{p}_x \rangle \langle  \hat{G} \rangle \big)=0=Q_{yz}\\
  Q_{xy}&=4 \big(\langle \hat{p}_x \hat{p}_y\rangle - \langle \hat{p}_x \rangle \langle \hat{p}_y \rangle\big)=0\\
\end{aligned}
\end{equation}

The re-expression of the result in matrix form is given by
\begin{equation}
   \begin{aligned}
       \mathcal{Q}_{l,p}=\left[\begin{array}{ccc}
\frac{4(2p+\vert l\vert+1)}{w_0^2} & 0 & 0 \\
0 & \frac{4(2p+\vert l\vert +1)}{w_0^2} & 0 \\
0 & 0 & \frac{2p(p+\vert l \vert)+2p+\vert l \vert +1}{z_R^2}
\end{array}\right].
   \end{aligned} 
\end{equation}

In the case of rotation modes, we assume an equal superposition of two LG modes with $l\neq l^{\prime}$, and $p=p^{\prime}=0$:

\begin{equation}
    \left|\Psi_{ll^{\prime}}\right\rangle=\frac1{\sqrt{2}}(\left|\mathrm{LG}_{l0}\right\rangle+\left|\mathrm{LG}_{l^{\prime}0}\right\rangle)=\frac1{\sqrt{2}}(\left|n_+,0\right\rangle+\left|n_+^{\prime},0\right\rangle).
\end{equation}

Using the same approach as Eq.~\ref{qfix}, the QFIm for rotation modes is given by 

\begin{equation}
   \begin{aligned}
       \mathcal{Q}_{l,l^{\prime}}=\left[\begin{array}{ccc}
\frac{2(\vert l\vert+\vert l^{\prime}\vert+2)}{w_0^2} & 0 & 0 \\
0 & \frac{2(\vert l\vert+\vert l^{\prime}\vert+2)}{w_0^2} & 0 \\
0 & 0 & \frac{[4+2(|l|+|l'|)+(|l|-|l'|)^2]}{z_R^2}
\end{array}\right].
   \end{aligned} 
\end{equation}

We proceed to derive the CFIm for ideal intensity detection. In this case, the measurement probability distribution is represented by the normalized intensity distribution of the light field, as $I(x, y \vert \theta) = \vert \tilde{\Psi}_{(x_e,y_e,z_e)}\vert^2$. Since the detector has infinitesimal pixels, we can obtain the summation form of CFIm calculation through indefinite integration in the spatial domain, which is expressed as:

\begin{equation}
\label{intensity measurement}
    \mathcal{F}_{ij}\left(\rho_{\theta}, \boldsymbol{\Pi}_{x,y}\right)=\iint \frac{1}{I(x, y\lvert\theta)} \frac{\partial I(x, y \vert \theta)}{\partial \theta_{i}} \frac{\partial I(x, y \vert\theta)}{\partial \theta_{j}} \mathrm{d} x \mathrm{d}y.
\end{equation}

Changing the integration variable $t=2\left[(x-x_e)^2+(y-y_e)^2\right]/w(z_e)^2$ and carrying out the derivatives of Laguerre polynomials using the relation $\partial_tL_p^l(t)=-L_{p-1}^{l+1}(t)$, the diagonal elements of CFIm can be obtained in compact expressions:

\begin{equation}
\begin{aligned}
    &\mathcal{F}_{x_ex_e}=\frac{4 p!}{(\vert l\vert+p)!}\frac1{w(z_e)^2}\\
    &\times\int_t e^{-t}t^{\vert l\vert-1}\bigg[2tL_{p-1}^{\vert l\vert+1}(t)+(t-\vert l\vert)L_{p}^{\vert l\vert}(t)\bigg]^{2}dt=\mathcal{F}_{y_ey_e},\\
    &\mathcal{F}_{z_ez_e}=\frac{4p!}{(\vert l\vert+p)!}\bigg[\frac{\partial_{z_e}w(z_e)}{w(z_e)}\bigg]^2\\
    &\times\int_t e^{-t}t^{\vert l\vert}\bigg[2tL_{p-1}^{|l|+1}(t)+(t-|l|-1)L_p^{\vert l\vert}(t)\bigg]^2\mathrm{d}t.
\end{aligned}
\end{equation}

We first expand the complex integral term as the summation of  different Laguerre polynomials:

\begin{equation}
\begin{aligned}
&\bigg[2tL_{p-1}^{\vert l\vert+1}(t)+(t-\vert l\vert)L_{p}^{\vert l\vert}(t)\bigg]^{2}\\
&=l^2L_{p}^{l}(t)^2-2ltL_{p}^{L}(t)^2+t^2L_{p}^{l}(t)^2+4t^2L_{p-1}^{l+1}(t)^2\\
&-4ltL_{p}^{l}(t)L_{p-1}^{l+1}(t)+4t^2L_{p}^{l}(t)L_{p-1}^{l+1}(t)\\
&\bigg[2tL_{p-1}^{l+1}(t)+(t-l-1)L_{p}^{l}(t)\bigg]^{2}\\
&=4t^{2}L_{p-1}^{l+1}(t)^{2}+t^{2}L_{p}^{l}(t)^{2}-2(l+1)tL_{p}^{l}(t)^{2}\\
&+(l+1)^{2}L_{p}^{l}(t)^{2}+4t^{2}L_{p-1}^{l+1}(t)L_{p}^{l}(t)-4(l+1)tL_{p-1}^{l+1}(t)L_{p}^{l}(t)\\
\end{aligned}
\end{equation}
Each integral can be further expanded according to the orthogonal polynomial \cite{rassias1992orthogonality} as

\begin{equation}
\begin{aligned}
 &\int_t\mathrm{e}^{-t}t^\mu L_p^l(t)L_{p^{\prime}}^{l^{\prime}}(t)\mathrm{d}t=(-1)^{p+p^{\prime}}\Gamma(\mu+1)\\
 &\times\sum_{k=0}^{\min(p,p^{\prime})}\binom{\mu-l}{p-k}\binom{\mu-l^{\prime}}{p^{\prime}-k}\binom{\mu+k}k.
\end{aligned}
\end{equation}

The following properties are important for simplifying this summation equation:

\begin{equation}
    \begin{aligned}
      \Gamma(n)&=(n-1)\Gamma(n-1),\\
      \sum_{k=0}^{m}{\binom{n+k}k}&=\binom{n+m}m+\binom{n+m}{m-1}.
    \end{aligned}
\end{equation}

Complex equations can be simplified and given analytic results as
\begin{equation}
\begin{aligned}
 \mathcal{F}_{x_ex_e}&=\frac{4}{{w(z)}^{2}}(2p+1)=\mathcal{F}_{y_ey_e},\\
  \mathcal{F}_{z_ez_e}&=\frac{4}{{R(z)}^{2}}[2p(p+\vert l \vert)+2p+\vert l \vert +1].
\end{aligned}
\end{equation}

The same routine can be applied to the non-diagonal entries of the matrix, yielding a value of zero. The CFIm of ideal intensity detection for pure LG modes is given by:

\begin{equation}
     \mathcal{F}\left(\rho_{\theta}, \boldsymbol{\Pi}_{x,y}\right)=4\left[\begin{array}{ccc}
\frac{(2p+1)}{{w(z)}^{2}} & 0 & 0 \\
0 & \frac{(2p+1)}{{w(z)}^{2}} & 0 \\
0 & 0 & \frac{2p(p+\vert l \vert)+2p+\vert l \vert +1}{{R(z)}^{2}}\end{array}\right].
\end{equation}

We have derived the analytical forms for the QFIm and CFIm as discussed in the main text. However, obtaining the analytical form for the rotational modes proved challenging. To facilitate quantitative analysis, we present numerical results for the rotational modes. Notably, the non-diagonal entries in the $x$ and $y$ coordinate matrix are non-zero, indicating mutual influence between the lateral coordinates due to the absence of circular symmetry of the intensity distribution. Consequently, the lateral localization precisions exhibit slight directional differences.

\section{Analyzing Model Mismatch}
\label{model mismatch}

Model mismatches can arise from various factors, including aberrations in the imaging system or limitations in point spread function (PSF) engineering techniques. The deviation between the observed image and the expected image is commonly quantified using the peak signal-to-noise ratio (PSNR) metric \cite{Accuracy2012, Scope2008}. Assessing axial aberrations involves examining the beam divergence, which can be evaluated using the M-square parameter \cite{Fundamentals2019}.

\begin{figure}[htbp]
\includegraphics[width=0.48\textwidth]{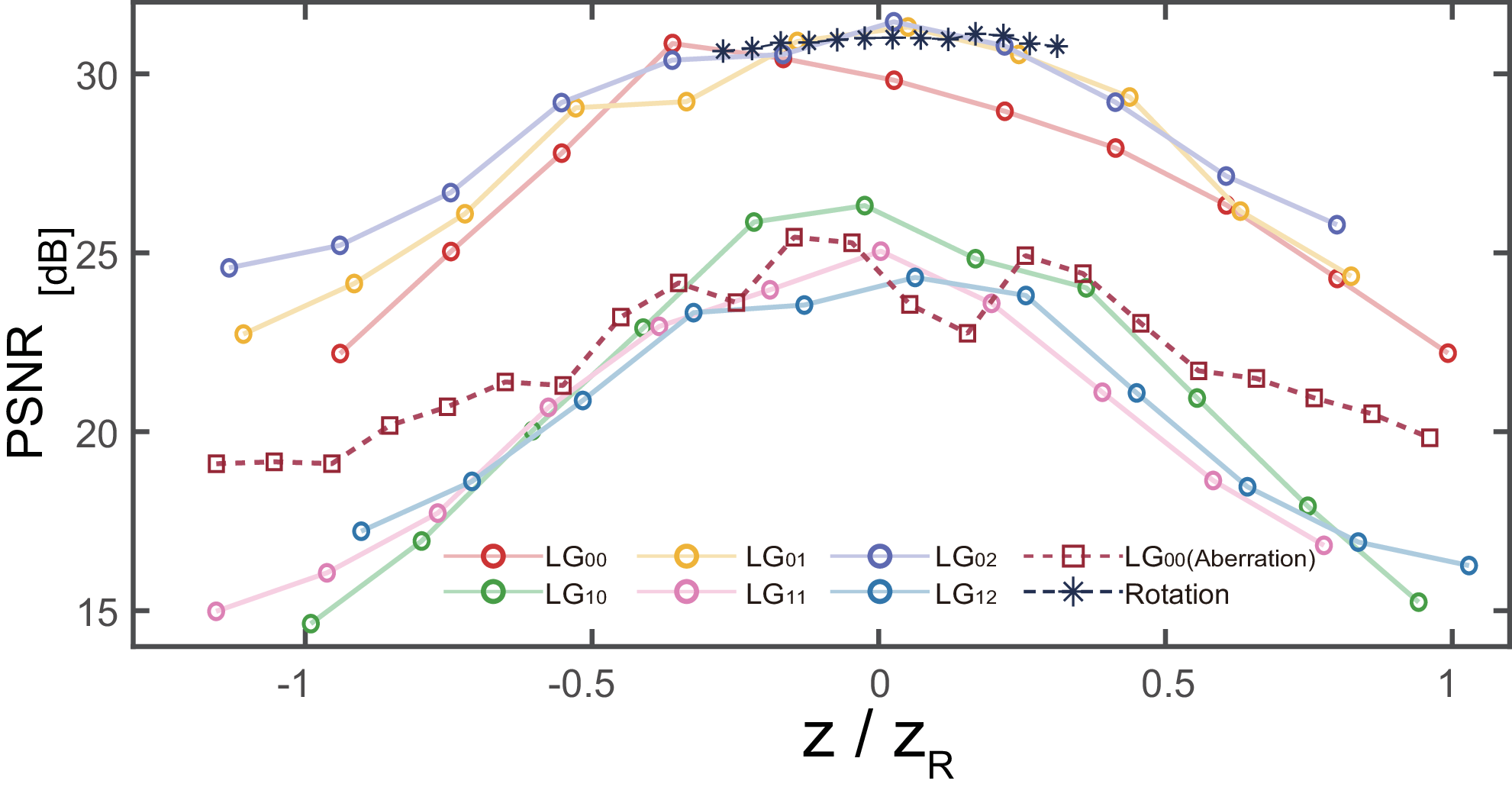}
\caption{\label{fig: psnr}  An estimation of beam lateral quality, as determined by the peak signal-to-noise ratio for the experimental intensity images.}
\end{figure}

To analyze the lateral mismatch, we employ the Matlab \textit{curve fitting tool} package for least squares (LS) fitting of the observed image to obtain the expected image. The raw data is averaged and processed with background subtraction and data normalization.  The MSE is calculated by comparing the experimental image $X$ of size $m \times n$ with the corresponding desired  reference image $K$:
\begin{equation}
MSE = \frac{1}{mn} \sum_{x=0}^{m-1} \sum_{y=0}^{n-1}[\boldsymbol{X}(x, y\vert\theta)-\boldsymbol{K}(x, y\vert \hat{\theta})]^2.
\end{equation}
The PSNR quality metric is then defined as:
\begin{equation}
PSNR = 10\log_{10}\frac{MAX_{K}^2}{MSE},
\end{equation}
where $MAX_{K}$ denotes the maximum value of $K$. The PSNR results are shown in Fig.~\ref{fig: psnr},  revealing high values in the vicinity of the focal plane.  

\begin{figure}[htbp]
\includegraphics[width=0.48\textwidth]{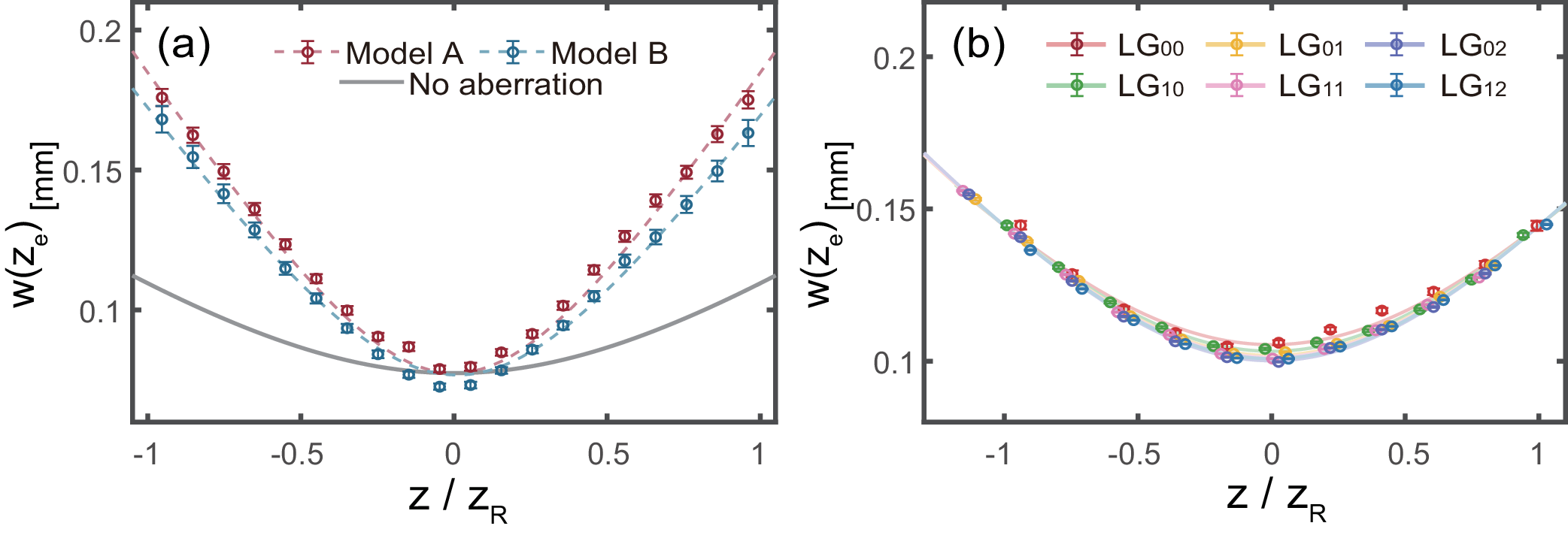}
\caption{\label{fig: waist} Experimental estimates of beam width for (a) imperfect Gaussian mode and for (b) LG modes at different defocusing distances. The theoretical results (solid lines) are calculated through $\hat{w}_0$ and the fitting results (dash lines) are obtained using LS fit.}
\end{figure}

In the case of generated LG modes, the PSNR diverges into two branches with $l = 0$ and $l = 1$, which stems from the fact that images with higher peak values generally yield higher PSNR values. Imperfect Gaussian modes, affected by aberrations, result in lower PSNR values compared to modulated Gaussian modes. Evaluation of axial aberrations is based on the M-square parameter, which represents the ratio of the experimental beam's angular divergence to the theoretical divergence, with the theoretical value given by $\mathcal{M}^2_{lp}=2p+l+1$ \cite{wan2022divergence}. The theoretical beam divergence is calculated using the estimated waist radius $\hat{w}0$, obtained through LS fitting of the experimentally estimated beam width $\hat{w}(z_e)$. The performance of the two algorithms differs not only in the axial variance but also in the mean value of the beam width obtained, as depicted in Fig. \ref{fig: waist}(a).  For the imperfect Gaussian mode with $l=0,p=0$, we obtained M-square values of approximately $\mathcal{M}^2_{00}\approx 1.42$ (Model A) and $\mathcal{M}^2_{00}\approx 1.52$ (Model B). These values indicate the presence of significant spherical aberration in our initial single-lens system. In contrast, modes generated through the 4-f system exhibit minimal spherical aberration, demonstrating excellent agreement with the theoretical divergences, as illustrated in Fig. \ref{fig: waist}(b).

\section{Comparison of Two Localization Algorithms}
\label{calculate NCRB}

For pixelated intensity detection, the elements of the CFIm can be obtained by summing the CFIs of each pixel:
\begin{equation}
\label{CFI pix}
  \mathcal{F}_{ij}(\rho_{\theta}, \boldsymbol{\Pi}_{A_k})=\sum_{k}\sum_{X_{k}} \frac{1}{P\left(X_{k} \vert\theta\right)}(\frac{\partial  P\left(X_{k} \vert\theta \right)}{\partial \theta_i}) (\frac{\partial  P\left(X_{k} \vert\theta \right)}{\partial \theta_j}).   
\end{equation}

In Model A, which falls under the category of mixed noise models involving two types of detector noise $N_b$ and $N_c$, the pixel readout probability distribution is given by \cite{Approaching2020}:
\begin{equation}
\label{PA}
\begin{aligned}
     P_\mathcal{A}(X_k \vert\mu_{k},\theta)&=\sum_{N_k}\mathcal{R}(X_k\vert N_k)P(N_k\vert\mu_{k},\theta),
\end{aligned}
\end{equation}
where $N_k$ represents the effective photon counts in $k$th pixel, which follows a Poisson distribution $P(N_k=\mu_k(x,y))$. The detector response, denoted by $\mathcal{R}(X_k\vert N_k)$, describes the relationship between the readout counts $X_k$ of the $k$th pixel and the effective photon count $N_k$. The response probability distribution is obtained through the convolution of the probability distributions of $N_b$ and $N_c$, which is given by
\begin{equation}
        \mathcal{R}(X_k\vert N_k)=\sum_{Nb}P(N_b)P(X_k-N_b\vert N_k).
\end{equation}

The elements of Model A's CFIm can be written as
\begin{equation}
    \begin{aligned}
&\mathcal{F_A}_{ij}(\theta) =\sum_{k}\sum_{X_{k}} \frac{1}{\sum_{N_k}\mathcal{R}(X_k\vert N_k)P(N_k\vert\mu_k,\theta)}\\
&\times \left[\sum_{N_{k}} \mathcal{R}(X_k\vert N_k) \frac{-e^{-\mu_k} \mu_k^{N_{k}}\frac{\partial \mu_k}{\partial \theta_i}+N_{k}e^{-\mu_k}\mu_k^{(N_{k}-1)}  \frac{\partial\mu_k}{\partial \theta_i}}{N_{k} !}\right] \\
&\times\left[\sum_{N_{k}} \mathcal{R}(X_k\vert N_k) \frac{-e^{-\mu_k} \mu_k^{N_{k}}\frac{\partial \mu_k}{\partial \theta_j}+N_{k}e^{-\mu_k}\mu_k^{(N_{k}-1)}  \frac{\partial\mu_k}{\partial \theta_j}}{N_{k} !}\right] \\
&=\sum_{k} \frac{1}{\mu_{k}(x, y)}\left(\frac{\partial \mu_{k}(x, y)}{\partial \theta_{i}}\right)\left(\frac{\partial \mu_{k}(x, y)}{\partial \theta_{j}}\right)\\ 
&\times\sum_{X_{k}} \frac{[\sum_{N_{k}} \mathcal{R}(X_k \vert N_k) P(N_{k} \vert \mu_k)(N_k-\mu_k)]^{2}}{\mu_k \sum_{N_k} \mathcal{R}(X_k \vert N_k) P(N_k \vert \mu_k)},
\end{aligned}
\end{equation}
which involve derivatives of the interest position parameters $\theta=({x_e,y_e,w(z_e)})$, and the corresponding pre-calibrated detector parameters $\theta=({N,N_b,N_c})$. The resulting CFIm is a 3x3 matrix given by
\begin{equation}
     \mathcal{F}_\mathcal{A}=\left[\begin{array}{ccc}
F_{x_e,x_e} & 0 & 0 \\
0 & F_{{y_e,y_e}} & 0 \\
0 & 0 & F_{w_e,w_e}\end{array}\right].
\end{equation}

With a known beam waist $w_0$, our program can be configured to directly measure the axial position $z$ instead of the beam radius $w$. This adjustment can be achieved through the substitution of the partial derivatives associated with the beam radius $w$ within the corresponding matrices, with those pertaining to the axial position $z$.

In Model B, only involving Poisson noise source $N_b$, the pixel readout probability distribution can be expressed in a simplified form, given by \cite{Fast2010}:
\begin{equation}
\label{PB}
    P_\mathcal{B}(X_k \vert\mu_{k},\theta)=\frac{\mu_{k}(x, y)^{X_{k}} e^{-\mu_{k}(x, y)}}{X_{k}!}.
\end{equation}
Utilizing the Stirling approximation ($\ln n ! \approx n \ln n-n$ for large $n$), the elements of Model B's CFIM can be simplified as follows:
\begin{equation}
   \mathcal{F_B}_{ij}(\theta)=\sum_{k} \frac{1}{\mu_{k}(x, y)} \frac{\partial \mu_{k}(x, y)}{\partial \theta_{i}} \frac{\partial \mu_{k}(x, y)}{\partial \theta_{j}},  
\end{equation}
which involve derivatives of five unknown parameters $\theta=({x_e,y_e,w(z_e),N, N_b})$. The resulting CFIm is a 5x5 matrix with a non-diagonal sub-matrix, given by
\begin{equation}
     \mathcal{F}_\mathcal{B}=\left[\begin{array}{ccccc}
F_{x,x} & 0 & 0 & 0 & 0\\
0 & F_{y,y} & 0 & 0 & 0 \\
0 & 0 & F_{w,w} & F_{w,N} & F_{w,Nb}\\
0 & 0 & F_{w,N} & F_{N,N} & F_{N,Nb}\\
0 & 0 & F_{w,Nb} & F_{N,Nb} & F_{Nb,Nb}\end{array}\right].
\end{equation}

The parameters $N$ and $N_b$ can be considered as nuisance parameters, potentially introducing imprecision in waist estimation. In the case of imperfect Gaussian mode localization experiments, our calibration results indicate that the average effective photon counts is $N_\mathcal{A}=11066$.  Conversely, direct estimation using Model B yields $N_\mathcal{B}=9782$ and $N_b\sim P(516.37)$, which are biased estimates. The results indicate that the variance of $N_b$ in Model B is exaggerated when approximating it as Poisson noise. Consequently, in the presence of aberrations and model mismatch, this bias affects the estimation of the beam width, as illustrated in Fig. \ref{fig: waist}(a).

\section{Effect of Pixelation and Noise}

\begin{figure}[htbp]
\includegraphics[width=0.48\textwidth]{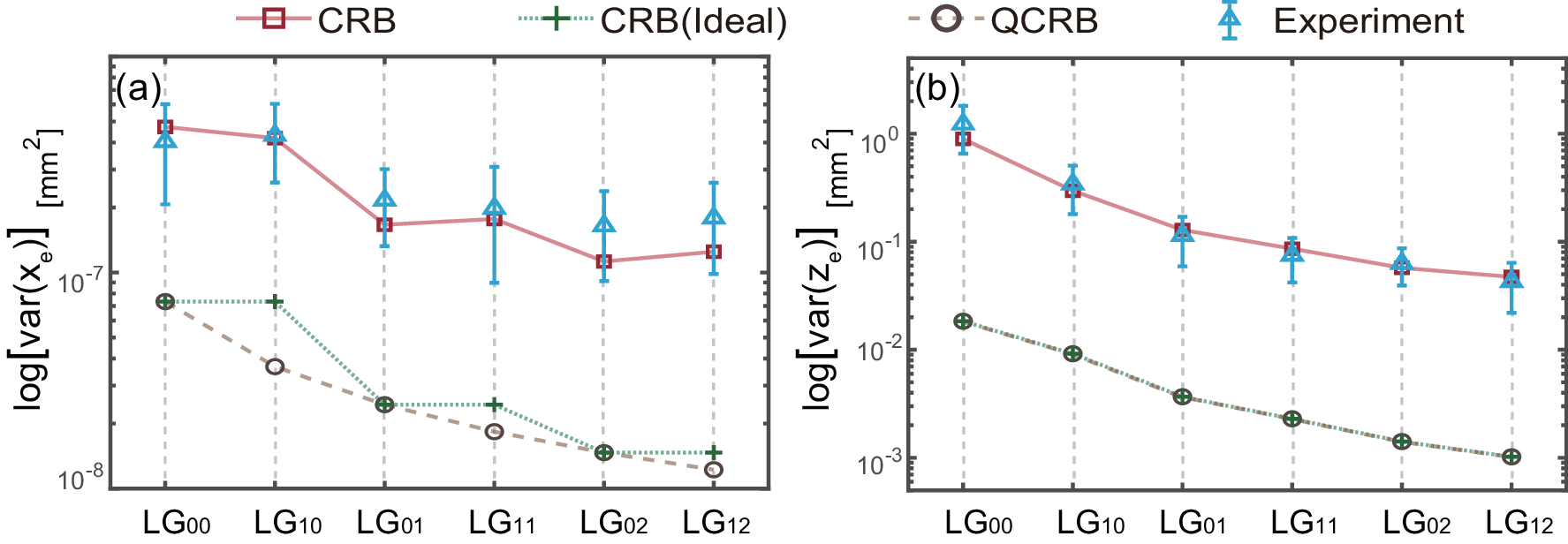}
\caption{\label{fig: best} Comparison of the (a) lateral and (b) axial localization precision of different modes in the optimal detection plane.}
\end{figure}

Pixelation and detection noise have a significant impact on localization precision and the determination of the optimal detection plane.  In the case of lateral localization, the optimal detection plane coincides with the beam waist, which exhibits the highest SNR. Conversely, for axial localization, the optimal position is at the Rayleigh range, where the SNR is relatively poor. Thus, there exists a trade-off between the SNR and the optimal position, shifting the optimal detection plane towards the waist.  The discrepancies between the practical CRB and the ideal CRB exhibit a greater prominence in the axial direction (Fig. \ref{fig: best}(b)) as opposed to the lateral direction (Fig. \ref{fig: best}(a)), owing to the higher SNR observed at the lateral optimal plane compared to the axial optimal plane.

\begin{figure*}[htbp]
\includegraphics[width=1\textwidth]{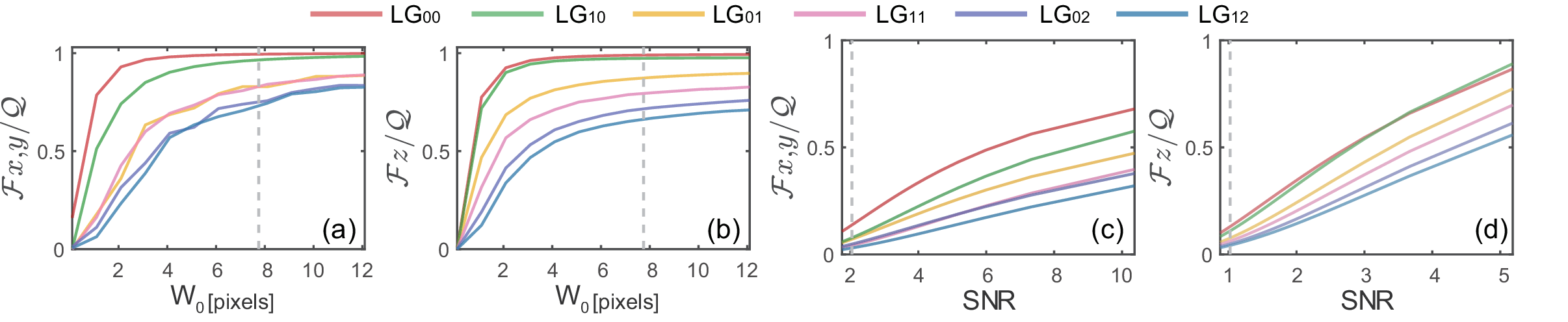}
\caption{\label{fig: noise} (a) the ratio between lateral CFI and QFI and (b) the ratio between axial CFI and QFI as functions of the waist-to-pixels ratio.  (c) the ratio between lateral CFI and QFI and (b) the ratio between axial CFI and QFI as functions of the SNR.  The dashed lines mark the waist-to-pixels ratio and the signal-to-noise ratio employed in the experiment.}
\end{figure*}

To further investigate the individual effects of pixelation and detector noise, we explore the variation of the ratio of CFI to QFI with increasing waist size in the absence of noise, as illustrated in Fig.~\ref{fig: noise}(a)-(b). It can be observed that the impact of pixelation becomes less significant as the spot size exceeds a certain threshold. Additionally, by employing the waist size utilized in our experimental setup, we analyze the influence of the detector noise in Fig.~\ref{fig: noise}(c)-(d). The ratio of CFI to QFI can be approximated as linearly dependent on the SNR. In contrast to pixelation, detector noise has a more substantial effect on the degradation of localization precision. The limited dynamic range of our detector restricts the achievable SNR in the experiments, resulting in the discrepancies in Fig. \ref{fig: best}. These discrepancies can be reduced significantly by improving SNR.

The impact of pixelation and noise is more pronounced in higher-order modes. This behavior can be attributed to the narrower central wave packet and the presence of additional side flaps in higher-order modes, rendering them more susceptible to the effects of pixelation. Moreover, the larger patterns associated with higher-order modes also lead to a reduction in the SNR of each pixel. Hence, when dealing with higher-order modes and more sophisticated superposition modes, it becomes even more crucial to conduct a thorough analysis of the actual accuracy based on a refined noise model.



\bibliography{output.bib}

\begin{thebibliography}{63}%
\makeatletter
\providecommand \@ifxundefined [1]{%
 \@ifx{#1\undefined}
}%
\providecommand \@ifnum [1]{%
 \ifnum #1\expandafter \@firstoftwo
 \else \expandafter \@secondoftwo
 \fi
}%
\providecommand \@ifx [1]{%
 \ifx #1\expandafter \@firstoftwo
 \else \expandafter \@secondoftwo
 \fi
}%
\providecommand \natexlab [1]{#1}%
\providecommand \enquote  [1]{``#1''}%
\providecommand \bibnamefont  [1]{#1}%
\providecommand \bibfnamefont [1]{#1}%
\providecommand \citenamefont [1]{#1}%
\providecommand \href@noop [0]{\@secondoftwo}%
\providecommand \href [0]{\begingroup \@sanitize@url \@href}%
\providecommand \@href[1]{\@@startlink{#1}\@@href}%
\providecommand \@@href[1]{\endgroup#1\@@endlink}%
\providecommand \@sanitize@url [0]{\catcode `\\12\catcode `\$12\catcode
  `\&12\catcode `\#12\catcode `\^12\catcode `\_12\catcode `\%12\relax}%
\providecommand \@@startlink[1]{}%
\providecommand \@@endlink[0]{}%
\providecommand \url  [0]{\begingroup\@sanitize@url \@url }%
\providecommand \@url [1]{\endgroup\@href {#1}{\urlprefix }}%
\providecommand \urlprefix  [0]{URL }%
\providecommand \Eprint [0]{\href }%
\providecommand \doibase [0]{https://doi.org/}%
\providecommand \selectlanguage [0]{\@gobble}%
\providecommand \bibinfo  [0]{\@secondoftwo}%
\providecommand \bibfield  [0]{\@secondoftwo}%
\providecommand \translation [1]{[#1]}%
\providecommand \BibitemOpen [0]{}%
\providecommand \bibitemStop [0]{}%
\providecommand \bibitemNoStop [0]{.\EOS\space}%
\providecommand \EOS [0]{\spacefactor3000\relax}%
\providecommand \BibitemShut  [1]{\csname bibitem#1\endcsname}%
\let\auto@bib@innerbib\@empty
\bibitem [{\citenamefont {Chen}\ \emph {et~al.}(2015)\citenamefont {Chen},
  \citenamefont {Zou}, \citenamefont {Gong}, \citenamefont {Dong},
  \citenamefont {Guo},\ and\ \citenamefont {Sun}}]{chen2015subdiffraction}%
  \BibitemOpen
  \bibfield  {author} {\bibinfo {author} {\bibfnamefont {X.}~\bibnamefont
  {Chen}}, \bibinfo {author} {\bibfnamefont {C.}~\bibnamefont {Zou}}, \bibinfo
  {author} {\bibfnamefont {Z.}~\bibnamefont {Gong}}, \bibinfo {author}
  {\bibfnamefont {C.}~\bibnamefont {Dong}}, \bibinfo {author} {\bibfnamefont
  {G.}~\bibnamefont {Guo}},\ and\ \bibinfo {author} {\bibfnamefont
  {F.}~\bibnamefont {Sun}},\ }\bibfield  {title} {\bibinfo {title}
  {Subdiffraction optical manipulation of the charge state of nitrogen vacancy
  center in diamond},\ }\href {https://doi.org/10.1038/lsa.2015.3} {\bibfield
  {journal} {\bibinfo  {journal} {Light: Science and Applications}\ }\textbf
  {\bibinfo {volume} {4}},\ \bibinfo {pages} {e230} (\bibinfo {year}
  {2015})}\BibitemShut {NoStop}%
\bibitem [{\citenamefont {Jaskula}\ \emph {et~al.}(2017)\citenamefont
  {Jaskula}, \citenamefont {Bauch}, \citenamefont {Arroyo-Camejo},
  \citenamefont {Lukin}, \citenamefont {Hell}, \citenamefont {Trifonov},\ and\
  \citenamefont {Walsworth}}]{jaskula2017superresolution}%
  \BibitemOpen
  \bibfield  {author} {\bibinfo {author} {\bibfnamefont {J.-C.}\ \bibnamefont
  {Jaskula}}, \bibinfo {author} {\bibfnamefont {E.}~\bibnamefont {Bauch}},
  \bibinfo {author} {\bibfnamefont {S.}~\bibnamefont {Arroyo-Camejo}}, \bibinfo
  {author} {\bibfnamefont {M.~D.}\ \bibnamefont {Lukin}}, \bibinfo {author}
  {\bibfnamefont {S.~W.}\ \bibnamefont {Hell}}, \bibinfo {author}
  {\bibfnamefont {A.~S.}\ \bibnamefont {Trifonov}},\ and\ \bibinfo {author}
  {\bibfnamefont {R.~L.}\ \bibnamefont {Walsworth}},\ }\bibfield  {title}
  {\bibinfo {title} {Superresolution optical magnetic imaging and spectroscopy
  using individual electronic spins in diamond},\ }\href
  {https://doi.org/10.1364/OE.25.011048} {\bibfield  {journal} {\bibinfo
  {journal} {Optics express}\ }\textbf {\bibinfo {volume} {25}},\ \bibinfo
  {pages} {11048} (\bibinfo {year} {2017})}\BibitemShut {NoStop}%
\bibitem [{\citenamefont {Dalgarno}\ \emph {et~al.}(2010)\citenamefont
  {Dalgarno}, \citenamefont {Dalgarno}, \citenamefont {Putoud}, \citenamefont
  {Lambert}, \citenamefont {Paterson}, \citenamefont {Logan}, \citenamefont
  {Towers}, \citenamefont {Warburton},\ and\ \citenamefont
  {Greenaway}}]{dalgarno2010multiplane}%
  \BibitemOpen
  \bibfield  {author} {\bibinfo {author} {\bibfnamefont {P.~A.}\ \bibnamefont
  {Dalgarno}}, \bibinfo {author} {\bibfnamefont {H.~I.}\ \bibnamefont
  {Dalgarno}}, \bibinfo {author} {\bibfnamefont {A.}~\bibnamefont {Putoud}},
  \bibinfo {author} {\bibfnamefont {R.}~\bibnamefont {Lambert}}, \bibinfo
  {author} {\bibfnamefont {L.}~\bibnamefont {Paterson}}, \bibinfo {author}
  {\bibfnamefont {D.~C.}\ \bibnamefont {Logan}}, \bibinfo {author}
  {\bibfnamefont {D.~P.}\ \bibnamefont {Towers}}, \bibinfo {author}
  {\bibfnamefont {R.~J.}\ \bibnamefont {Warburton}},\ and\ \bibinfo {author}
  {\bibfnamefont {A.~H.}\ \bibnamefont {Greenaway}},\ }\bibfield  {title}
  {\bibinfo {title} {Multiplane imaging and three dimensional nanoscale
  particle tracking in biological microscopy},\ }\href
  {https://doi.org/10.1364/OE.18.000877} {\bibfield  {journal} {\bibinfo
  {journal} {Optics express}\ }\textbf {\bibinfo {volume} {18}},\ \bibinfo
  {pages} {877} (\bibinfo {year} {2010})}\BibitemShut {NoStop}%
\bibitem [{\citenamefont {Abrahamsson}\ \emph {et~al.}(2013)\citenamefont
  {Abrahamsson}, \citenamefont {Chen}, \citenamefont {Hajj}, \citenamefont
  {Stallinga}, \citenamefont {Katsov}, \citenamefont {Wisniewski},
  \citenamefont {Mizuguchi}, \citenamefont {Soule}, \citenamefont {Mueller},
  \citenamefont {Darzacq} \emph {et~al.}}]{abrahamsson2013fast}%
  \BibitemOpen
  \bibfield  {author} {\bibinfo {author} {\bibfnamefont {S.}~\bibnamefont
  {Abrahamsson}}, \bibinfo {author} {\bibfnamefont {J.}~\bibnamefont {Chen}},
  \bibinfo {author} {\bibfnamefont {B.}~\bibnamefont {Hajj}}, \bibinfo {author}
  {\bibfnamefont {S.}~\bibnamefont {Stallinga}}, \bibinfo {author}
  {\bibfnamefont {A.~Y.}\ \bibnamefont {Katsov}}, \bibinfo {author}
  {\bibfnamefont {J.}~\bibnamefont {Wisniewski}}, \bibinfo {author}
  {\bibfnamefont {G.}~\bibnamefont {Mizuguchi}}, \bibinfo {author}
  {\bibfnamefont {P.}~\bibnamefont {Soule}}, \bibinfo {author} {\bibfnamefont
  {F.}~\bibnamefont {Mueller}}, \bibinfo {author} {\bibfnamefont {C.~D.}\
  \bibnamefont {Darzacq}}, \emph {et~al.},\ }\bibfield  {title} {\bibinfo
  {title} {Fast multicolor 3d imaging using aberration-corrected multifocus
  microscopy},\ }\href {https://doi.org/10.1038/NMETH.2277} {\bibfield
  {journal} {\bibinfo  {journal} {Nature methods}\ }\textbf {\bibinfo {volume}
  {10}},\ \bibinfo {pages} {60} (\bibinfo {year} {2013})}\BibitemShut {NoStop}%
\bibitem [{\citenamefont {Manzo}\ and\ \citenamefont
  {Garcia-Parajo}(2015)}]{manzo2015review}%
  \BibitemOpen
  \bibfield  {author} {\bibinfo {author} {\bibfnamefont {C.}~\bibnamefont
  {Manzo}}\ and\ \bibinfo {author} {\bibfnamefont {M.~F.}\ \bibnamefont
  {Garcia-Parajo}},\ }\bibfield  {title} {\bibinfo {title} {A review of
  progress in single particle tracking: from methods to biophysical insights},\
  }\href {https://doi.org/10.1088/0034-4885/78/12/124601} {\bibfield  {journal}
  {\bibinfo  {journal} {Reports on progress in physics}\ }\textbf {\bibinfo
  {volume} {78}},\ \bibinfo {pages} {124601} (\bibinfo {year}
  {2015})}\BibitemShut {NoStop}%
\bibitem [{\citenamefont {von Diezmann}\ \emph {et~al.}(2017)\citenamefont {von
  Diezmann}, \citenamefont {Shechtman},\ and\ \citenamefont
  {Moerner}}]{Three2017}%
  \BibitemOpen
  \bibfield  {author} {\bibinfo {author} {\bibfnamefont {L.}~\bibnamefont {von
  Diezmann}}, \bibinfo {author} {\bibfnamefont {Y.}~\bibnamefont {Shechtman}},\
  and\ \bibinfo {author} {\bibfnamefont {W.~E.}\ \bibnamefont {Moerner}},\
  }\bibfield  {title} {\bibinfo {title} {Three-{Dimensional} {Localization} of
  {Single} {Molecules} for {Super}-{Resolution} {Imaging} and
  {Single}-{Particle} {Tracking}},\ }\href
  {https://doi.org/10.1021/acs.chemrev.6b00629} {\bibfield  {journal} {\bibinfo
   {journal} {Chem. Rev.}\ }\textbf {\bibinfo {volume} {117}},\ \bibinfo
  {pages} {7244} (\bibinfo {year} {2017})}\BibitemShut {NoStop}%
\bibitem [{\citenamefont {Shen}\ \emph {et~al.}(2017)\citenamefont {Shen},
  \citenamefont {Tauzin}, \citenamefont {Baiyasi}, \citenamefont {Wang},
  \citenamefont {Moringo}, \citenamefont {Shuang},\ and\ \citenamefont
  {Landes}}]{shen2017single}%
  \BibitemOpen
  \bibfield  {author} {\bibinfo {author} {\bibfnamefont {H.}~\bibnamefont
  {Shen}}, \bibinfo {author} {\bibfnamefont {L.~J.}\ \bibnamefont {Tauzin}},
  \bibinfo {author} {\bibfnamefont {R.}~\bibnamefont {Baiyasi}}, \bibinfo
  {author} {\bibfnamefont {W.}~\bibnamefont {Wang}}, \bibinfo {author}
  {\bibfnamefont {N.}~\bibnamefont {Moringo}}, \bibinfo {author} {\bibfnamefont
  {B.}~\bibnamefont {Shuang}},\ and\ \bibinfo {author} {\bibfnamefont {C.~F.}\
  \bibnamefont {Landes}},\ }\bibfield  {title} {\bibinfo {title} {Single
  particle tracking: from theory to biophysical applications},\ }\href
  {https://doi.org/10.1021/acs.chemrev.6b00815} {\bibfield  {journal} {\bibinfo
   {journal} {Chemical reviews}\ }\textbf {\bibinfo {volume} {117}},\ \bibinfo
  {pages} {7331} (\bibinfo {year} {2017})}\BibitemShut {NoStop}%
\bibitem [{\citenamefont {Hell}\ and\ \citenamefont
  {Wichmann}(1994)}]{Breaking1994}%
  \BibitemOpen
  \bibfield  {author} {\bibinfo {author} {\bibfnamefont {S.~W.}\ \bibnamefont
  {Hell}}\ and\ \bibinfo {author} {\bibfnamefont {J.}~\bibnamefont
  {Wichmann}},\ }\bibfield  {title} {\bibinfo {title} {Breaking the diffraction
  resolution limit by stimulated emission: stimulated-emission-depletion
  fluorescence microscopy},\ }\href {https://doi.org/10.1364/OL.19.000780}
  {\bibfield  {journal} {\bibinfo  {journal} {Opt. Lett.}\ }\textbf {\bibinfo
  {volume} {19}},\ \bibinfo {pages} {780} (\bibinfo {year} {1994})}\BibitemShut
  {NoStop}%
\bibitem [{\citenamefont {Betzig}\ \emph {et~al.}(2006)\citenamefont {Betzig},
  \citenamefont {Patterson}, \citenamefont {Sougrat}, \citenamefont
  {Lindwasser}, \citenamefont {Olenych}, \citenamefont {Bonifacino},
  \citenamefont {Davidson}, \citenamefont {Lippincott-Schwartz},\ and\
  \citenamefont {Hess}}]{imaging2006}%
  \BibitemOpen
  \bibfield  {author} {\bibinfo {author} {\bibfnamefont {E.}~\bibnamefont
  {Betzig}}, \bibinfo {author} {\bibfnamefont {G.~H.}\ \bibnamefont
  {Patterson}}, \bibinfo {author} {\bibfnamefont {R.}~\bibnamefont {Sougrat}},
  \bibinfo {author} {\bibfnamefont {O.~W.}\ \bibnamefont {Lindwasser}},
  \bibinfo {author} {\bibfnamefont {S.}~\bibnamefont {Olenych}}, \bibinfo
  {author} {\bibfnamefont {J.~S.}\ \bibnamefont {Bonifacino}}, \bibinfo
  {author} {\bibfnamefont {M.~W.}\ \bibnamefont {Davidson}}, \bibinfo {author}
  {\bibfnamefont {J.}~\bibnamefont {Lippincott-Schwartz}},\ and\ \bibinfo
  {author} {\bibfnamefont {H.~F.}\ \bibnamefont {Hess}},\ }\bibfield  {title}
  {\bibinfo {title} {Imaging intracellular fluorescent proteins at nanometer
  resolution},\ }\href {https://doi.org/10.1126/science.1127344} {\bibfield
  {journal} {\bibinfo  {journal} {science}\ }\textbf {\bibinfo {volume}
  {313}},\ \bibinfo {pages} {1642} (\bibinfo {year} {2006})}\BibitemShut
  {NoStop}%
\bibitem [{\citenamefont {Rust}\ \emph {et~al.}(2006)\citenamefont {Rust},
  \citenamefont {Bates},\ and\ \citenamefont {Zhuang}}]{rust2006stochastic}%
  \BibitemOpen
  \bibfield  {author} {\bibinfo {author} {\bibfnamefont {M.~J.}\ \bibnamefont
  {Rust}}, \bibinfo {author} {\bibfnamefont {M.}~\bibnamefont {Bates}},\ and\
  \bibinfo {author} {\bibfnamefont {X.}~\bibnamefont {Zhuang}},\ }\bibfield
  {title} {\bibinfo {title} {Stochastic optical reconstruction microscopy
  (storm) provides sub-diffraction-limit image resolution},\ }\href
  {https://doi.org/10.1038/nmeth929} {\bibfield  {journal} {\bibinfo  {journal}
  {Nature methods}\ }\textbf {\bibinfo {volume} {3}},\ \bibinfo {pages} {793}
  (\bibinfo {year} {2006})}\BibitemShut {NoStop}%
\bibitem [{\citenamefont {Huang}\ \emph {et~al.}(2008)\citenamefont {Huang},
  \citenamefont {Wang}, \citenamefont {Bates},\ and\ \citenamefont
  {Zhuang}}]{Three2008}%
  \BibitemOpen
  \bibfield  {author} {\bibinfo {author} {\bibfnamefont {B.}~\bibnamefont
  {Huang}}, \bibinfo {author} {\bibfnamefont {W.}~\bibnamefont {Wang}},
  \bibinfo {author} {\bibfnamefont {M.}~\bibnamefont {Bates}},\ and\ \bibinfo
  {author} {\bibfnamefont {X.}~\bibnamefont {Zhuang}},\ }\bibfield  {title}
  {\bibinfo {title} {Three-dimensional super-resolution imaging by stochastic
  optical reconstruction microscopy},\ }\href
  {https://doi.org/10.1126/science.1153529} {\bibfield  {journal} {\bibinfo
  {journal} {Science}\ }\textbf {\bibinfo {volume} {319}},\ \bibinfo {pages}
  {810} (\bibinfo {year} {2008})}\BibitemShut {NoStop}%
\bibitem [{\citenamefont {Rayleigh}(1879)}]{XXXI1879}%
  \BibitemOpen
  \bibfield  {author} {\bibinfo {author} {\bibfnamefont {L.}~\bibnamefont
  {Rayleigh}},\ }\bibfield  {title} {\bibinfo {title} {Xxxi. investigations in
  optics, with special reference to the spectroscope},\ }\href@noop {}
  {\bibfield  {journal} {\bibinfo  {journal} {Philos. Mag.}\ }\textbf {\bibinfo
  {volume} {8}},\ \bibinfo {pages} {261} (\bibinfo {year} {1879})}\BibitemShut
  {NoStop}%
\bibitem [{\citenamefont {Born}\ and\ \citenamefont
  {Wolf}(1999)}]{Principles1999}%
  \BibitemOpen
  \bibfield  {author} {\bibinfo {author} {\bibfnamefont {M.}~\bibnamefont
  {Born}}\ and\ \bibinfo {author} {\bibfnamefont {E.}~\bibnamefont {Wolf}},\
  }\bibfield  {title} {\bibinfo {title} {Principles of optics, 7th (expanded)
  edition},\ }\href@noop {} {\bibfield  {journal} {\bibinfo  {journal} {United
  Kingdom: Press Syndicate of the University of Cambridge}\ }\textbf {\bibinfo
  {volume} {461}},\ \bibinfo {pages} {401} (\bibinfo {year}
  {1999})}\BibitemShut {NoStop}%
\bibitem [{\citenamefont {Tsang}\ \emph {et~al.}(2016)\citenamefont {Tsang},
  \citenamefont {Nair},\ and\ \citenamefont {Lu}}]{tsang2016quantum}%
  \BibitemOpen
  \bibfield  {author} {\bibinfo {author} {\bibfnamefont {M.}~\bibnamefont
  {Tsang}}, \bibinfo {author} {\bibfnamefont {R.}~\bibnamefont {Nair}},\ and\
  \bibinfo {author} {\bibfnamefont {X.-M.}\ \bibnamefont {Lu}},\ }\bibfield
  {title} {\bibinfo {title} {Quantum theory of superresolution for two
  incoherent optical point sources},\ }\href
  {https://doi.org/10.1103/PhysRevX.6.031033} {\bibfield  {journal} {\bibinfo
  {journal} {Physical Review X}\ }\textbf {\bibinfo {volume} {6}},\ \bibinfo
  {pages} {031033} (\bibinfo {year} {2016})}\BibitemShut {NoStop}%
\bibitem [{\citenamefont {Tsang}(2017)}]{tsang2017subdiffraction}%
  \BibitemOpen
  \bibfield  {author} {\bibinfo {author} {\bibfnamefont {M.}~\bibnamefont
  {Tsang}},\ }\bibfield  {title} {\bibinfo {title} {Subdiffraction incoherent
  optical imaging via spatial-mode demultiplexing},\ }\href
  {https://doi.org/10.1088/1367-2630/aa5b9c} {\bibfield  {journal} {\bibinfo
  {journal} {New Journal of Physics}\ }\textbf {\bibinfo {volume} {19}},\
  \bibinfo {pages} {023054} (\bibinfo {year} {2017})}\BibitemShut {NoStop}%
\bibitem [{\citenamefont {Tsang}(2018)}]{tsang2018subdiffraction}%
  \BibitemOpen
  \bibfield  {author} {\bibinfo {author} {\bibfnamefont {M.}~\bibnamefont
  {Tsang}},\ }\bibfield  {title} {\bibinfo {title} {Subdiffraction incoherent
  optical imaging via spatial-mode demultiplexing: Semiclassical treatment},\
  }\href {https://doi.org/https://doi.org/10.1103/PhysRevA.97.023830}
  {\bibfield  {journal} {\bibinfo  {journal} {Physical Review A}\ }\textbf
  {\bibinfo {volume} {97}},\ \bibinfo {pages} {023830} (\bibinfo {year}
  {2018})}\BibitemShut {NoStop}%
\bibitem [{\citenamefont {Tsang}(2015)}]{Quantum2015}%
  \BibitemOpen
  \bibfield  {author} {\bibinfo {author} {\bibfnamefont {M.}~\bibnamefont
  {Tsang}},\ }\bibfield  {title} {\bibinfo {title} {Quantum limits to optical
  point-source localization},\ }\href {https://doi.org/10.1364/OPTICA.2.000646}
  {\bibfield  {journal} {\bibinfo  {journal} {Optica}\ }\textbf {\bibinfo
  {volume} {2}},\ \bibinfo {pages} {646} (\bibinfo {year} {2015})}\BibitemShut
  {NoStop}%
\bibitem [{\citenamefont {Backlund}\ \emph {et~al.}(2018)\citenamefont
  {Backlund}, \citenamefont {Shechtman},\ and\ \citenamefont
  {Walsworth}}]{Fundamental2018}%
  \BibitemOpen
  \bibfield  {author} {\bibinfo {author} {\bibfnamefont {M.~P.}\ \bibnamefont
  {Backlund}}, \bibinfo {author} {\bibfnamefont {Y.}~\bibnamefont
  {Shechtman}},\ and\ \bibinfo {author} {\bibfnamefont {R.~L.}\ \bibnamefont
  {Walsworth}},\ }\bibfield  {title} {\bibinfo {title} {Fundamental precision
  bounds for three-dimensional optical localization microscopy with {Poisson}
  statistics},\ }\href {https://doi.org/10.1103/PhysRevLett.121.023904}
  {\bibfield  {journal} {\bibinfo  {journal} {Phys. Rev. Lett.}\ }\textbf
  {\bibinfo {volume} {121}},\ \bibinfo {pages} {023904} (\bibinfo {year}
  {2018})},\ \bibinfo {note} {arXiv:1803.01776 [physics]}\BibitemShut {NoStop}%
\bibitem [{\citenamefont {Yu}\ and\ \citenamefont
  {Prasad}(2018)}]{yu2018quantum}%
  \BibitemOpen
  \bibfield  {author} {\bibinfo {author} {\bibfnamefont {Z.}~\bibnamefont
  {Yu}}\ and\ \bibinfo {author} {\bibfnamefont {S.}~\bibnamefont {Prasad}},\
  }\bibfield  {title} {\bibinfo {title} {Quantum limited superresolution of an
  incoherent source pair in three dimensions},\ }\href
  {https://doi.org/10.1103/PhysRevLett.121.180504} {\bibfield  {journal}
  {\bibinfo  {journal} {Physical review letters}\ }\textbf {\bibinfo {volume}
  {121}},\ \bibinfo {pages} {180504} (\bibinfo {year} {2018})}\BibitemShut
  {NoStop}%
\bibitem [{\citenamefont {Helstrom}(1969)}]{Quantum1969}%
  \BibitemOpen
  \bibfield  {author} {\bibinfo {author} {\bibfnamefont {C.~W.}\ \bibnamefont
  {Helstrom}},\ }\href {https://link.springer.com/article/10.1007/BF01007479}
  {\emph {\bibinfo {title} {Quantum detection and estimation theory}}},\
  Vol.~\bibinfo {volume} {1}\ (\bibinfo  {publisher} {Springer},\ \bibinfo
  {year} {1969})\ pp.\ \bibinfo {pages} {231--252}\BibitemShut {NoStop}%
\bibitem [{\citenamefont {Braunstein}\ and\ \citenamefont
  {Caves}(1994)}]{Statistical1994}%
  \BibitemOpen
  \bibfield  {author} {\bibinfo {author} {\bibfnamefont {S.~L.}\ \bibnamefont
  {Braunstein}}\ and\ \bibinfo {author} {\bibfnamefont {C.~M.}\ \bibnamefont
  {Caves}},\ }\bibfield  {title} {\bibinfo {title} {Statistical distance and
  the geometry of quantum states},\ }\href
  {https://doi.org/https://doi.org/10.1103/PhysRevLett.72.3439} {\bibfield
  {journal} {\bibinfo  {journal} {Physical Review Letters}\ }\textbf {\bibinfo
  {volume} {72}},\ \bibinfo {pages} {3439} (\bibinfo {year}
  {1994})}\BibitemShut {NoStop}%
\bibitem [{\citenamefont {Yang}\ \emph {et~al.}(2016)\citenamefont {Yang},
  \citenamefont {Tashchilina}, \citenamefont {Moiseev}, \citenamefont {Simon},\
  and\ \citenamefont {Lvovsky}}]{yang2016far}%
  \BibitemOpen
  \bibfield  {author} {\bibinfo {author} {\bibfnamefont {F.}~\bibnamefont
  {Yang}}, \bibinfo {author} {\bibfnamefont {A.}~\bibnamefont {Tashchilina}},
  \bibinfo {author} {\bibfnamefont {E.~S.}\ \bibnamefont {Moiseev}}, \bibinfo
  {author} {\bibfnamefont {C.}~\bibnamefont {Simon}},\ and\ \bibinfo {author}
  {\bibfnamefont {A.~I.}\ \bibnamefont {Lvovsky}},\ }\bibfield  {title}
  {\bibinfo {title} {Far-field linear optical superresolution via heterodyne
  detection in a higher-order local oscillator mode},\ }\href
  {https://doi.org/10.1364/OPTICA.3.001148} {\bibfield  {journal} {\bibinfo
  {journal} {Optica}\ }\textbf {\bibinfo {volume} {3}},\ \bibinfo {pages}
  {1148} (\bibinfo {year} {2016})}\BibitemShut {NoStop}%
\bibitem [{\citenamefont {Tang}\ \emph {et~al.}(2016)\citenamefont {Tang},
  \citenamefont {Durak},\ and\ \citenamefont {Ling}}]{tang2016fault}%
  \BibitemOpen
  \bibfield  {author} {\bibinfo {author} {\bibfnamefont {Z.~S.}\ \bibnamefont
  {Tang}}, \bibinfo {author} {\bibfnamefont {K.}~\bibnamefont {Durak}},\ and\
  \bibinfo {author} {\bibfnamefont {A.}~\bibnamefont {Ling}},\ }\bibfield
  {title} {\bibinfo {title} {Fault-tolerant and finite-error localization for
  point emitters within the diffraction limit},\ }\href
  {https://doi.org/10.1364/OE.24.022004} {\bibfield  {journal} {\bibinfo
  {journal} {Optics express}\ }\textbf {\bibinfo {volume} {24}},\ \bibinfo
  {pages} {22004} (\bibinfo {year} {2016})}\BibitemShut {NoStop}%
\bibitem [{\citenamefont {Dutton}\ \emph {et~al.}(2019)\citenamefont {Dutton},
  \citenamefont {Kerviche}, \citenamefont {Ashok},\ and\ \citenamefont
  {Guha}}]{dutton2019attaining}%
  \BibitemOpen
  \bibfield  {author} {\bibinfo {author} {\bibfnamefont {Z.}~\bibnamefont
  {Dutton}}, \bibinfo {author} {\bibfnamefont {R.}~\bibnamefont {Kerviche}},
  \bibinfo {author} {\bibfnamefont {A.}~\bibnamefont {Ashok}},\ and\ \bibinfo
  {author} {\bibfnamefont {S.}~\bibnamefont {Guha}},\ }\bibfield  {title}
  {\bibinfo {title} {Attaining the quantum limit of superresolution in imaging
  an object's length via predetection spatial-mode sorting},\ }\href
  {https://doi.org/10.1103/PhysRevA.99.033847} {\bibfield  {journal} {\bibinfo
  {journal} {Physical Review A}\ }\textbf {\bibinfo {volume} {99}},\ \bibinfo
  {pages} {033847} (\bibinfo {year} {2019})}\BibitemShut {NoStop}%
\bibitem [{\citenamefont {Zhou}\ \emph {et~al.}(2019)\citenamefont {Zhou},
  \citenamefont {Yang}, \citenamefont {Hassett}, \citenamefont {Rafsanjani},
  \citenamefont {Mirhosseini}, \citenamefont {Vamivakas}, \citenamefont
  {Jordan}, \citenamefont {Shi},\ and\ \citenamefont {Boyd}}]{zhou2019quantum}%
  \BibitemOpen
  \bibfield  {author} {\bibinfo {author} {\bibfnamefont {Y.}~\bibnamefont
  {Zhou}}, \bibinfo {author} {\bibfnamefont {J.}~\bibnamefont {Yang}}, \bibinfo
  {author} {\bibfnamefont {J.~D.}\ \bibnamefont {Hassett}}, \bibinfo {author}
  {\bibfnamefont {S.~M.~H.}\ \bibnamefont {Rafsanjani}}, \bibinfo {author}
  {\bibfnamefont {M.}~\bibnamefont {Mirhosseini}}, \bibinfo {author}
  {\bibfnamefont {A.~N.}\ \bibnamefont {Vamivakas}}, \bibinfo {author}
  {\bibfnamefont {A.~N.}\ \bibnamefont {Jordan}}, \bibinfo {author}
  {\bibfnamefont {Z.}~\bibnamefont {Shi}},\ and\ \bibinfo {author}
  {\bibfnamefont {R.~W.}\ \bibnamefont {Boyd}},\ }\bibfield  {title} {\bibinfo
  {title} {Quantum-limited estimation of the axial separation of two incoherent
  point sources},\ }\href {https://doi.org/10.1364/OPTICA.6.000534} {\bibfield
  {journal} {\bibinfo  {journal} {Optica}\ }\textbf {\bibinfo {volume} {6}},\
  \bibinfo {pages} {534} (\bibinfo {year} {2019})}\BibitemShut {NoStop}%
\bibitem [{\citenamefont {Liu}\ \emph {et~al.}(2020)\citenamefont {Liu},
  \citenamefont {Yuan}, \citenamefont {Lu},\ and\ \citenamefont
  {Wang}}]{liu2020quantum}%
  \BibitemOpen
  \bibfield  {author} {\bibinfo {author} {\bibfnamefont {J.}~\bibnamefont
  {Liu}}, \bibinfo {author} {\bibfnamefont {H.}~\bibnamefont {Yuan}}, \bibinfo
  {author} {\bibfnamefont {X.-M.}\ \bibnamefont {Lu}},\ and\ \bibinfo {author}
  {\bibfnamefont {X.}~\bibnamefont {Wang}},\ }\bibfield  {title} {\bibinfo
  {title} {Quantum fisher information matrix and multiparameter estimation},\
  }\href@noop {} {\bibfield  {journal} {\bibinfo  {journal} {Journal of Physics
  A: Mathematical and Theoretical}\ }\textbf {\bibinfo {volume} {53}},\
  \bibinfo {pages} {023001} (\bibinfo {year} {2020})}\BibitemShut {NoStop}%
\bibitem [{\citenamefont {Albarelli}\ \emph {et~al.}(2020)\citenamefont
  {Albarelli}, \citenamefont {Barbieri}, \citenamefont {Genoni},\ and\
  \citenamefont {Gianani}}]{albarelli2020perspective}%
  \BibitemOpen
  \bibfield  {author} {\bibinfo {author} {\bibfnamefont {F.}~\bibnamefont
  {Albarelli}}, \bibinfo {author} {\bibfnamefont {M.}~\bibnamefont {Barbieri}},
  \bibinfo {author} {\bibfnamefont {M.~G.}\ \bibnamefont {Genoni}},\ and\
  \bibinfo {author} {\bibfnamefont {I.}~\bibnamefont {Gianani}},\ }\bibfield
  {title} {\bibinfo {title} {A perspective on multiparameter quantum metrology:
  From theoretical tools to applications in quantum imaging},\ }\href
  {https://doi.org/10.1016/j.physleta.2020.126311} {\bibfield  {journal}
  {\bibinfo  {journal} {Physics Letters A}\ }\textbf {\bibinfo {volume}
  {384}},\ \bibinfo {pages} {126311} (\bibinfo {year} {2020})}\BibitemShut
  {NoStop}%
\bibitem [{\citenamefont {Holtzer}\ \emph {et~al.}(2007)\citenamefont
  {Holtzer}, \citenamefont {Meckel},\ and\ \citenamefont
  {Schmidt}}]{Nanometric2007}%
  \BibitemOpen
  \bibfield  {author} {\bibinfo {author} {\bibfnamefont {L.}~\bibnamefont
  {Holtzer}}, \bibinfo {author} {\bibfnamefont {T.}~\bibnamefont {Meckel}},\
  and\ \bibinfo {author} {\bibfnamefont {T.}~\bibnamefont {Schmidt}},\
  }\bibfield  {title} {\bibinfo {title} {Nanometric three-dimensional tracking
  of individual quantum dots in cells},\ }\href
  {https://doi.org/https://doi.org/10.1063/1.2437066} {\bibfield  {journal}
  {\bibinfo  {journal} {Applied Physics Letters}\ }\textbf {\bibinfo {volume}
  {90}},\ \bibinfo {pages} {053902} (\bibinfo {year} {2007})}\BibitemShut
  {NoStop}%
\bibitem [{\citenamefont {Pavani}\ and\ \citenamefont
  {Piestun}(2008)}]{High2008}%
  \BibitemOpen
  \bibfield  {author} {\bibinfo {author} {\bibfnamefont {S.~R.~P.}\
  \bibnamefont {Pavani}}\ and\ \bibinfo {author} {\bibfnamefont
  {R.}~\bibnamefont {Piestun}},\ }\bibfield  {title} {\bibinfo {title}
  {High-efficiency rotating point spread functions},\ }\href
  {https://doi.org/10.1364/OE.16.003484} {\bibfield  {journal} {\bibinfo
  {journal} {Optics express}\ }\textbf {\bibinfo {volume} {16}},\ \bibinfo
  {pages} {3484} (\bibinfo {year} {2008})}\BibitemShut {NoStop}%
\bibitem [{\citenamefont {Pavani}\ \emph {et~al.}(2009)\citenamefont {Pavani},
  \citenamefont {Thompson}, \citenamefont {Biteen}, \citenamefont {Lord},
  \citenamefont {Liu}, \citenamefont {Twieg}, \citenamefont {Piestun},\ and\
  \citenamefont {Moerner}}]{Three2009}%
  \BibitemOpen
  \bibfield  {author} {\bibinfo {author} {\bibfnamefont {S.~R.~P.}\
  \bibnamefont {Pavani}}, \bibinfo {author} {\bibfnamefont {M.~A.}\
  \bibnamefont {Thompson}}, \bibinfo {author} {\bibfnamefont {J.~S.}\
  \bibnamefont {Biteen}}, \bibinfo {author} {\bibfnamefont {S.~J.}\
  \bibnamefont {Lord}}, \bibinfo {author} {\bibfnamefont {N.}~\bibnamefont
  {Liu}}, \bibinfo {author} {\bibfnamefont {R.~J.}\ \bibnamefont {Twieg}},
  \bibinfo {author} {\bibfnamefont {R.}~\bibnamefont {Piestun}},\ and\ \bibinfo
  {author} {\bibfnamefont {W.~E.}\ \bibnamefont {Moerner}},\ }\bibfield
  {title} {\bibinfo {title} {Three-dimensional, single-molecule fluorescence
  imaging beyond the diffraction limit by using a double-helix point spread
  function},\ }\href@noop {} {\bibfield  {journal} {\bibinfo  {journal}
  {Proceedings of the National Academy of Sciences}\ }\textbf {\bibinfo
  {volume} {106}},\ \bibinfo {pages} {2995} (\bibinfo {year}
  {2009})}\BibitemShut {NoStop}%
\bibitem [{\citenamefont {Wang}\ \emph {et~al.}(2021)\citenamefont {Wang},
  \citenamefont {Xu}, \citenamefont {Li},\ and\ \citenamefont
  {Zhang}}]{Quantum2021}%
  \BibitemOpen
  \bibfield  {author} {\bibinfo {author} {\bibfnamefont {B.}~\bibnamefont
  {Wang}}, \bibinfo {author} {\bibfnamefont {L.}~\bibnamefont {Xu}}, \bibinfo
  {author} {\bibfnamefont {J.-c.}\ \bibnamefont {Li}},\ and\ \bibinfo {author}
  {\bibfnamefont {L.}~\bibnamefont {Zhang}},\ }\bibfield  {title} {\bibinfo
  {title} {Quantum-limited localization and resolution in three dimensions},\
  }\href {https://doi.org/10.1364/PRJ.417613} {\bibfield  {journal} {\bibinfo
  {journal} {Photon. Res.}\ }\textbf {\bibinfo {volume} {9}},\ \bibinfo {pages}
  {1522} (\bibinfo {year} {2021})}\BibitemShut {NoStop}%
\bibitem [{\citenamefont {{\v{R}}eh{\'a}{\v{c}}ek}\ \emph
  {et~al.}(2019)\citenamefont {{\v{R}}eh{\'a}{\v{c}}ek}, \citenamefont
  {Pa{\'u}r}, \citenamefont {Stoklasa}, \citenamefont {Koutn{\`y}},
  \citenamefont {Hradil},\ and\ \citenamefont
  {S{\'a}nchez-Soto}}]{Intensity2019}%
  \BibitemOpen
  \bibfield  {author} {\bibinfo {author} {\bibfnamefont {J.}~\bibnamefont
  {{\v{R}}eh{\'a}{\v{c}}ek}}, \bibinfo {author} {\bibfnamefont
  {M.}~\bibnamefont {Pa{\'u}r}}, \bibinfo {author} {\bibfnamefont
  {B.}~\bibnamefont {Stoklasa}}, \bibinfo {author} {\bibfnamefont
  {D.}~\bibnamefont {Koutn{\`y}}}, \bibinfo {author} {\bibfnamefont
  {Z.}~\bibnamefont {Hradil}},\ and\ \bibinfo {author} {\bibfnamefont
  {L.}~\bibnamefont {S{\'a}nchez-Soto}},\ }\bibfield  {title} {\bibinfo {title}
  {Intensity-based axial localization at the quantum limit},\ }\href
  {https://doi.org/10.1103/PhysRevLett.123.193601} {\bibfield  {journal}
  {\bibinfo  {journal} {Physical Review Letters}\ }\textbf {\bibinfo {volume}
  {123}},\ \bibinfo {pages} {193601} (\bibinfo {year} {2019})}\BibitemShut
  {NoStop}%
\bibitem [{\citenamefont {Koutn{\`y}}\ \emph {et~al.}(2021)\citenamefont
  {Koutn{\`y}}, \citenamefont {Hradil}, \citenamefont
  {{\v{R}}eh{\'a}{\v{c}}ek},\ and\ \citenamefont
  {S{\'a}nchez-Soto}}]{Axial2021}%
  \BibitemOpen
  \bibfield  {author} {\bibinfo {author} {\bibfnamefont {D.}~\bibnamefont
  {Koutn{\`y}}}, \bibinfo {author} {\bibfnamefont {Z.}~\bibnamefont {Hradil}},
  \bibinfo {author} {\bibfnamefont {J.}~\bibnamefont
  {{\v{R}}eh{\'a}{\v{c}}ek}},\ and\ \bibinfo {author} {\bibfnamefont
  {L.}~\bibnamefont {S{\'a}nchez-Soto}},\ }\bibfield  {title} {\bibinfo {title}
  {Axial superlocalization with vortex beams},\ }\href
  {https://doi.org/10.1088/2058-9565/abe8ca} {\bibfield  {journal} {\bibinfo
  {journal} {Quantum Science and Technology}\ }\textbf {\bibinfo {volume}
  {6}},\ \bibinfo {pages} {025021} (\bibinfo {year} {2021})}\BibitemShut
  {NoStop}%
\bibitem [{\citenamefont {Linowski}\ \emph {et~al.}(2023)\citenamefont
  {Linowski}, \citenamefont {Schlichtholz}, \citenamefont {Sorelli},
  \citenamefont {Gessner}, \citenamefont {Walschaers}, \citenamefont {Treps},\
  and\ \citenamefont {Rudnicki}}]{linowski2023application}%
  \BibitemOpen
  \bibfield  {author} {\bibinfo {author} {\bibfnamefont {T.}~\bibnamefont
  {Linowski}}, \bibinfo {author} {\bibfnamefont {K.}~\bibnamefont
  {Schlichtholz}}, \bibinfo {author} {\bibfnamefont {G.}~\bibnamefont
  {Sorelli}}, \bibinfo {author} {\bibfnamefont {M.}~\bibnamefont {Gessner}},
  \bibinfo {author} {\bibfnamefont {M.}~\bibnamefont {Walschaers}}, \bibinfo
  {author} {\bibfnamefont {N.}~\bibnamefont {Treps}},\ and\ \bibinfo {author}
  {\bibfnamefont {{\L}.}~\bibnamefont {Rudnicki}},\ }\bibfield  {title}
  {\bibinfo {title} {Application range of crosstalk-affected spatial
  demultiplexing for resolving separations between unbalanced sources},\ }\href
  {https://doi.org/10.1088/1367-2630/ad0173} {\bibfield  {journal} {\bibinfo
  {journal} {New Journal of Physics}\ }\textbf {\bibinfo {volume} {25}},\
  \bibinfo {pages} {103050} (\bibinfo {year} {2023})}\BibitemShut {NoStop}%
\bibitem [{\citenamefont {Matsumoto}(2002)}]{New2002}%
  \BibitemOpen
  \bibfield  {author} {\bibinfo {author} {\bibfnamefont {K.}~\bibnamefont
  {Matsumoto}},\ }\bibfield  {title} {\bibinfo {title} {A new approach to the
  cram{\'e}r-rao-type bound of the pure-state model},\ }\href@noop {}
  {\bibfield  {journal} {\bibinfo  {journal} {Journal of Physics A:
  Mathematical and General}\ }\textbf {\bibinfo {volume} {35}},\ \bibinfo
  {pages} {3111} (\bibinfo {year} {2002})}\BibitemShut {NoStop}%
\bibitem [{\citenamefont {Ragy}\ \emph {et~al.}(2016)\citenamefont {Ragy},
  \citenamefont {Jarzyna},\ and\ \citenamefont
  {Demkowicz-Dobrza{\'n}ski}}]{Compatibility2016}%
  \BibitemOpen
  \bibfield  {author} {\bibinfo {author} {\bibfnamefont {S.}~\bibnamefont
  {Ragy}}, \bibinfo {author} {\bibfnamefont {M.}~\bibnamefont {Jarzyna}},\ and\
  \bibinfo {author} {\bibfnamefont {R.}~\bibnamefont
  {Demkowicz-Dobrza{\'n}ski}},\ }\bibfield  {title} {\bibinfo {title}
  {Compatibility in multiparameter quantum metrology},\ }\href
  {https://doi.org/10.1103/PhysRevA.99.029905} {\bibfield  {journal} {\bibinfo
  {journal} {Physical Review A}\ }\textbf {\bibinfo {volume} {94}},\ \bibinfo
  {pages} {052108} (\bibinfo {year} {2016})}\BibitemShut {NoStop}%
\bibitem [{\citenamefont {Saleh}\ and\ \citenamefont
  {Teich}(2019)}]{Fundamentals2019}%
  \BibitemOpen
  \bibfield  {author} {\bibinfo {author} {\bibfnamefont {B.~E.}\ \bibnamefont
  {Saleh}}\ and\ \bibinfo {author} {\bibfnamefont {M.~C.}\ \bibnamefont
  {Teich}},\ }\href
  {https://books.google.com.hk/books?hl=zh-CN&lr=&id=rcqKDwAAQBAJ&oi=fnd&pg=PR1&dq=Fundamentals+of+photonics&ots=tHklg6zDwY&sig=ZBpku9YJCvJOnI2atOtL92wNI4g&redir_esc=y#v=onepage&q=Fundamentals%20of%20photonics&f=false}
  {\emph {\bibinfo {title} {Fundamentals of photonics}}},\ \bibinfo {number}
  {Sec 3.1, Sec 4.4}\ (\bibinfo  {publisher} {john Wiley \& sons},\ \bibinfo
  {year} {2019})\BibitemShut {NoStop}%
\bibitem [{\citenamefont {Goodman}(2005)}]{Introduction2005}%
  \BibitemOpen
  \bibfield  {author} {\bibinfo {author} {\bibfnamefont {J.~W.}\ \bibnamefont
  {Goodman}},\ }\href
  {https://books.google.com.hk/books?hl=zh-CN&lr=&id=ow5xs_Rtt9AC&oi=fnd&pg=PR7&dq=Introduction+to+Fourier+optics&ots=GZp9BO4ELH&sig=wgQ7V9B__Wxx2qjqKKN7gF9FvSQ&redir_esc=y#v=onepage&q=Introduction%20to%20Fourier%20optics&f=false}
  {\emph {\bibinfo {title} {Introduction to Fourier optics}}}\ (\bibinfo
  {publisher} {Roberts and Company publishers},\ \bibinfo {year}
  {2005})\BibitemShut {NoStop}%
\bibitem [{\citenamefont {Schechner}\ \emph {et~al.}(1996)\citenamefont
  {Schechner}, \citenamefont {Piestun},\ and\ \citenamefont
  {Shamir}}]{Wave1996}%
  \BibitemOpen
  \bibfield  {author} {\bibinfo {author} {\bibfnamefont {Y.~Y.}\ \bibnamefont
  {Schechner}}, \bibinfo {author} {\bibfnamefont {R.}~\bibnamefont {Piestun}},\
  and\ \bibinfo {author} {\bibfnamefont {J.}~\bibnamefont {Shamir}},\
  }\bibfield  {title} {\bibinfo {title} {Wave propagation with rotating
  intensity distributions},\ }\href {https://doi.org/10.1103/PhysRevE.54.R50}
  {\bibfield  {journal} {\bibinfo  {journal} {Physical Review E}\ }\textbf
  {\bibinfo {volume} {54}},\ \bibinfo {pages} {R50} (\bibinfo {year}
  {1996})}\BibitemShut {NoStop}%
\bibitem [{\citenamefont {Piestun}\ \emph {et~al.}(2000)\citenamefont
  {Piestun}, \citenamefont {Schechner},\ and\ \citenamefont
  {Shamir}}]{Propagation2000}%
  \BibitemOpen
  \bibfield  {author} {\bibinfo {author} {\bibfnamefont {R.}~\bibnamefont
  {Piestun}}, \bibinfo {author} {\bibfnamefont {Y.~Y.}\ \bibnamefont
  {Schechner}},\ and\ \bibinfo {author} {\bibfnamefont {J.}~\bibnamefont
  {Shamir}},\ }\bibfield  {title} {\bibinfo {title} {Propagation-invariant wave
  fields with finite energy},\ }\href {https://doi.org/10.1364/JOSAA.17.000294}
  {\bibfield  {journal} {\bibinfo  {journal} {JOSA A}\ }\textbf {\bibinfo
  {volume} {17}},\ \bibinfo {pages} {294} (\bibinfo {year} {2000})}\BibitemShut
  {NoStop}%
\bibitem [{\citenamefont {Shechtman}\ \emph {et~al.}(2014)\citenamefont
  {Shechtman}, \citenamefont {Sahl}, \citenamefont {Backer},\ and\
  \citenamefont {Moerner}}]{Optimal2014}%
  \BibitemOpen
  \bibfield  {author} {\bibinfo {author} {\bibfnamefont {Y.}~\bibnamefont
  {Shechtman}}, \bibinfo {author} {\bibfnamefont {S.~J.}\ \bibnamefont {Sahl}},
  \bibinfo {author} {\bibfnamefont {A.~S.}\ \bibnamefont {Backer}},\ and\
  \bibinfo {author} {\bibfnamefont {W.~E.}\ \bibnamefont {Moerner}},\
  }\bibfield  {title} {\bibinfo {title} {Optimal point spread function design
  for 3d imaging},\ }\href {https://doi.org/10.1103/PhysRevLett.113.133902}
  {\bibfield  {journal} {\bibinfo  {journal} {Physical review letters}\
  }\textbf {\bibinfo {volume} {113}},\ \bibinfo {pages} {133902} (\bibinfo
  {year} {2014})}\BibitemShut {NoStop}%
\bibitem [{\citenamefont {Greengard}\ \emph {et~al.}(2006)\citenamefont
  {Greengard}, \citenamefont {Schechner},\ and\ \citenamefont
  {Piestun}}]{Depth2006}%
  \BibitemOpen
  \bibfield  {author} {\bibinfo {author} {\bibfnamefont {A.}~\bibnamefont
  {Greengard}}, \bibinfo {author} {\bibfnamefont {Y.~Y.}\ \bibnamefont
  {Schechner}},\ and\ \bibinfo {author} {\bibfnamefont {R.}~\bibnamefont
  {Piestun}},\ }\bibfield  {title} {\bibinfo {title} {Depth from diffracted
  rotation},\ }\href {https://doi.org/10.1364/OL.31.000181} {\bibfield
  {journal} {\bibinfo  {journal} {Optics letters}\ }\textbf {\bibinfo {volume}
  {31}},\ \bibinfo {pages} {181} (\bibinfo {year} {2006})}\BibitemShut
  {NoStop}%
\bibitem [{\citenamefont {Wan}\ \emph {et~al.}(2022)\citenamefont {Wan},
  \citenamefont {Shen}, \citenamefont {Wang}, \citenamefont {Shi},
  \citenamefont {Liu},\ and\ \citenamefont {Fu}}]{wan2022divergence}%
  \BibitemOpen
  \bibfield  {author} {\bibinfo {author} {\bibfnamefont {Z.}~\bibnamefont
  {Wan}}, \bibinfo {author} {\bibfnamefont {Y.}~\bibnamefont {Shen}}, \bibinfo
  {author} {\bibfnamefont {Z.}~\bibnamefont {Wang}}, \bibinfo {author}
  {\bibfnamefont {Z.}~\bibnamefont {Shi}}, \bibinfo {author} {\bibfnamefont
  {Q.}~\bibnamefont {Liu}},\ and\ \bibinfo {author} {\bibfnamefont
  {X.}~\bibnamefont {Fu}},\ }\bibfield  {title} {\bibinfo {title}
  {Divergence-degenerate spatial multiplexing towards future ultrahigh
  capacity, low error-rate optical communications},\ }\href
  {https://doi.org/10.1038/s41377-022-00834-4} {\bibfield  {journal} {\bibinfo
  {journal} {Light: Science \& Applications}\ }\textbf {\bibinfo {volume}
  {11}},\ \bibinfo {pages} {144} (\bibinfo {year} {2022})}\BibitemShut
  {NoStop}%
\bibitem [{\citenamefont {Ober}\ \emph {et~al.}(2004)\citenamefont {Ober},
  \citenamefont {Ram},\ and\ \citenamefont {Ward}}]{Localization2004}%
  \BibitemOpen
  \bibfield  {author} {\bibinfo {author} {\bibfnamefont {R.~J.}\ \bibnamefont
  {Ober}}, \bibinfo {author} {\bibfnamefont {S.}~\bibnamefont {Ram}},\ and\
  \bibinfo {author} {\bibfnamefont {E.~S.}\ \bibnamefont {Ward}},\ }\bibfield
  {title} {\bibinfo {title} {Localization accuracy in single-molecule
  microscopy},\ }\href {https://doi.org/10.1016/S0006-3495(04)74193-4}
  {\bibfield  {journal} {\bibinfo  {journal} {Biophysical journal}\ }\textbf
  {\bibinfo {volume} {86}},\ \bibinfo {pages} {1185} (\bibinfo {year}
  {2004})}\BibitemShut {NoStop}%
\bibitem [{\citenamefont {Ram}\ \emph {et~al.}(2006)\citenamefont {Ram},
  \citenamefont {Sally~Ward},\ and\ \citenamefont {Ober}}]{Stochastic2006}%
  \BibitemOpen
  \bibfield  {author} {\bibinfo {author} {\bibfnamefont {S.}~\bibnamefont
  {Ram}}, \bibinfo {author} {\bibfnamefont {E.}~\bibnamefont {Sally~Ward}},\
  and\ \bibinfo {author} {\bibfnamefont {R.~J.}\ \bibnamefont {Ober}},\
  }\bibfield  {title} {\bibinfo {title} {A {Stochastic} {Analysis} of
  {Performance} {Limits} for {Optical} {Microscopes}},\ }\href
  {https://doi.org/10.1007/s11045-005-6237-2} {\bibfield  {journal} {\bibinfo
  {journal} {Multidim Syst Sign Process}\ }\textbf {\bibinfo {volume} {17}},\
  \bibinfo {pages} {27} (\bibinfo {year} {2006})}\BibitemShut {NoStop}%
\bibitem [{\citenamefont {Smith}\ \emph {et~al.}(2010)\citenamefont {Smith},
  \citenamefont {Joseph}, \citenamefont {Rieger},\ and\ \citenamefont
  {Lidke}}]{Fast2010}%
  \BibitemOpen
  \bibfield  {author} {\bibinfo {author} {\bibfnamefont {C.~S.}\ \bibnamefont
  {Smith}}, \bibinfo {author} {\bibfnamefont {N.}~\bibnamefont {Joseph}},
  \bibinfo {author} {\bibfnamefont {B.}~\bibnamefont {Rieger}},\ and\ \bibinfo
  {author} {\bibfnamefont {K.~A.}\ \bibnamefont {Lidke}},\ }\bibfield  {title}
  {\bibinfo {title} {Fast, single-molecule localization that achieves
  theoretically minimum uncertainty},\ }\href
  {https://doi.org/10.1038/NMETH.1449} {\bibfield  {journal} {\bibinfo
  {journal} {Nature methods}\ }\textbf {\bibinfo {volume} {7}},\ \bibinfo
  {pages} {373} (\bibinfo {year} {2010})}\BibitemShut {NoStop}%
\bibitem [{\citenamefont {Abraham}\ \emph {et~al.}(2009)\citenamefont
  {Abraham}, \citenamefont {Ram}, \citenamefont {Chao}, \citenamefont {Ward},\
  and\ \citenamefont {Ober}}]{Quantitative2009}%
  \BibitemOpen
  \bibfield  {author} {\bibinfo {author} {\bibfnamefont {A.~V.}\ \bibnamefont
  {Abraham}}, \bibinfo {author} {\bibfnamefont {S.}~\bibnamefont {Ram}},
  \bibinfo {author} {\bibfnamefont {J.}~\bibnamefont {Chao}}, \bibinfo {author}
  {\bibfnamefont {E.}~\bibnamefont {Ward}},\ and\ \bibinfo {author}
  {\bibfnamefont {R.~J.}\ \bibnamefont {Ober}},\ }\bibfield  {title} {\bibinfo
  {title} {Quantitative study of single molecule location estimation
  techniques},\ }\href {https://doi.org/10.1364/OE.17.023352} {\bibfield
  {journal} {\bibinfo  {journal} {Optics express}\ }\textbf {\bibinfo {volume}
  {17}},\ \bibinfo {pages} {23352} (\bibinfo {year} {2009})}\BibitemShut
  {NoStop}%
\bibitem [{\citenamefont {Janesick}\ \emph {et~al.}(1987)\citenamefont
  {Janesick}, \citenamefont {Elliott}, \citenamefont {Collins}, \citenamefont
  {Blouke},\ and\ \citenamefont {Freeman}}]{janesick1987scientific}%
  \BibitemOpen
  \bibfield  {author} {\bibinfo {author} {\bibfnamefont {J.~R.}\ \bibnamefont
  {Janesick}}, \bibinfo {author} {\bibfnamefont {T.}~\bibnamefont {Elliott}},
  \bibinfo {author} {\bibfnamefont {S.}~\bibnamefont {Collins}}, \bibinfo
  {author} {\bibfnamefont {M.~M.}\ \bibnamefont {Blouke}},\ and\ \bibinfo
  {author} {\bibfnamefont {J.}~\bibnamefont {Freeman}},\ }\bibfield  {title}
  {\bibinfo {title} {Scientific charge-coupled devices},\ }\href
  {https://doi.org/https://doi.org/10.1117/12.7974139} {\bibfield  {journal}
  {\bibinfo  {journal} {Optical Engineering}\ }\textbf {\bibinfo {volume}
  {26}},\ \bibinfo {pages} {692} (\bibinfo {year} {1987})}\BibitemShut
  {NoStop}%
\bibitem [{\citenamefont {Xu}\ \emph {et~al.}(2020)\citenamefont {Xu},
  \citenamefont {Liu}, \citenamefont {Datta}, \citenamefont {Knee},
  \citenamefont {Lundeen}, \citenamefont {Lu},\ and\ \citenamefont
  {Zhang}}]{Approaching2020}%
  \BibitemOpen
  \bibfield  {author} {\bibinfo {author} {\bibfnamefont {L.}~\bibnamefont
  {Xu}}, \bibinfo {author} {\bibfnamefont {Z.}~\bibnamefont {Liu}}, \bibinfo
  {author} {\bibfnamefont {A.}~\bibnamefont {Datta}}, \bibinfo {author}
  {\bibfnamefont {G.~C.}\ \bibnamefont {Knee}}, \bibinfo {author}
  {\bibfnamefont {J.~S.}\ \bibnamefont {Lundeen}}, \bibinfo {author}
  {\bibfnamefont {Y.-q.}\ \bibnamefont {Lu}},\ and\ \bibinfo {author}
  {\bibfnamefont {L.}~\bibnamefont {Zhang}},\ }\bibfield  {title} {\bibinfo
  {title} {Approaching {Quantum}-{Limited} {Metrology} with {Imperfect}
  {Detectors} by {Using} {Weak}-{Value} {Amplification}},\ }\href
  {https://doi.org/10.1103/PhysRevLett.125.080501} {\bibfield  {journal}
  {\bibinfo  {journal} {Phys. Rev. Lett.}\ }\textbf {\bibinfo {volume} {125}},\
  \bibinfo {pages} {080501} (\bibinfo {year} {2020})}\BibitemShut {NoStop}%
\bibitem [{\citenamefont {Yin}\ \emph {et~al.}(2021)\citenamefont {Yin},
  \citenamefont {Zhang}, \citenamefont {Xu}, \citenamefont {Liu}, \citenamefont
  {Zhuang}, \citenamefont {Chen}, \citenamefont {Gong}, \citenamefont {Ma},
  \citenamefont {Peng}, \citenamefont {Li} \emph {et~al.}}]{Improving2021}%
  \BibitemOpen
  \bibfield  {author} {\bibinfo {author} {\bibfnamefont {P.}~\bibnamefont
  {Yin}}, \bibinfo {author} {\bibfnamefont {W.-H.}\ \bibnamefont {Zhang}},
  \bibinfo {author} {\bibfnamefont {L.}~\bibnamefont {Xu}}, \bibinfo {author}
  {\bibfnamefont {Z.-G.}\ \bibnamefont {Liu}}, \bibinfo {author} {\bibfnamefont
  {W.-F.}\ \bibnamefont {Zhuang}}, \bibinfo {author} {\bibfnamefont
  {L.}~\bibnamefont {Chen}}, \bibinfo {author} {\bibfnamefont {M.}~\bibnamefont
  {Gong}}, \bibinfo {author} {\bibfnamefont {Y.}~\bibnamefont {Ma}}, \bibinfo
  {author} {\bibfnamefont {X.-X.}\ \bibnamefont {Peng}}, \bibinfo {author}
  {\bibfnamefont {G.-C.}\ \bibnamefont {Li}}, \emph {et~al.},\ }\bibfield
  {title} {\bibinfo {title} {Improving the precision of optical metrology by
  detecting fewer photons with biased weak measurement},\ }\href
  {https://doi.org/10.1038/s41377-021-00543-4} {\bibfield  {journal} {\bibinfo
  {journal} {Light: Science \& Applications}\ }\textbf {\bibinfo {volume}
  {10}},\ \bibinfo {pages} {103} (\bibinfo {year} {2021})}\BibitemShut
  {NoStop}%
\bibitem [{\citenamefont {Aguet}\ \emph {et~al.}(2005)\citenamefont {Aguet},
  \citenamefont {Van De~Ville},\ and\ \citenamefont
  {Unser}}]{aguet2005maximum}%
  \BibitemOpen
  \bibfield  {author} {\bibinfo {author} {\bibfnamefont {F.}~\bibnamefont
  {Aguet}}, \bibinfo {author} {\bibfnamefont {D.}~\bibnamefont {Van
  De~Ville}},\ and\ \bibinfo {author} {\bibfnamefont {M.}~\bibnamefont
  {Unser}},\ }\bibfield  {title} {\bibinfo {title} {A maximum-likelihood
  formalism for sub-resolution axial localization of fluorescent
  nanoparticles},\ }\href {https://doi.org/10.1364/OPEX.13.010503} {\bibfield
  {journal} {\bibinfo  {journal} {Optics Express}\ }\textbf {\bibinfo {volume}
  {13}},\ \bibinfo {pages} {10503} (\bibinfo {year} {2005})}\BibitemShut
  {NoStop}%
\bibitem [{\citenamefont {Kay}(1993)}]{Fundamentals1993}%
  \BibitemOpen
  \bibfield  {author} {\bibinfo {author} {\bibfnamefont {S.~M.}\ \bibnamefont
  {Kay}},\ }\href@noop {} {\emph {\bibinfo {title} {Fundamentals of statistical
  signal processing: estimation theory}}}\ (\bibinfo  {publisher}
  {Prentice-Hall, Inc.},\ \bibinfo {year} {1993})\BibitemShut {NoStop}%
\bibitem [{\citenamefont {Small}(2018)}]{Spherical2018}%
  \BibitemOpen
  \bibfield  {author} {\bibinfo {author} {\bibfnamefont {A.}~\bibnamefont
  {Small}},\ }\bibfield  {title} {\bibinfo {title} {Spherical aberration, coma,
  and the abbe sine condition for physicists who don't design lenses},\ }\href
  {https://doi.org/10.1119/1.5036939} {\bibfield  {journal} {\bibinfo
  {journal} {American Journal of Physics}\ }\textbf {\bibinfo {volume} {86}},\
  \bibinfo {pages} {487} (\bibinfo {year} {2018})}\BibitemShut {NoStop}%
\bibitem [{\citenamefont {Clark}\ \emph {et~al.}(2016)\citenamefont {Clark},
  \citenamefont {Offer}, \citenamefont {Franke-Arnold}, \citenamefont
  {Arnold},\ and\ \citenamefont {Radwell}}]{Comparison2016}%
  \BibitemOpen
  \bibfield  {author} {\bibinfo {author} {\bibfnamefont {T.~W.}\ \bibnamefont
  {Clark}}, \bibinfo {author} {\bibfnamefont {R.~F.}\ \bibnamefont {Offer}},
  \bibinfo {author} {\bibfnamefont {S.}~\bibnamefont {Franke-Arnold}}, \bibinfo
  {author} {\bibfnamefont {A.~S.}\ \bibnamefont {Arnold}},\ and\ \bibinfo
  {author} {\bibfnamefont {N.}~\bibnamefont {Radwell}},\ }\bibfield  {title}
  {\bibinfo {title} {Comparison of beam generation techniques using a phase
  only spatial light modulator},\ }\href {https://doi.org/10.1364/OE.24.006249}
  {\bibfield  {journal} {\bibinfo  {journal} {Optics express}\ }\textbf
  {\bibinfo {volume} {24}},\ \bibinfo {pages} {6249} (\bibinfo {year}
  {2016})}\BibitemShut {NoStop}%
\bibitem [{\citenamefont {Arriz{\'o}n}\ \emph {et~al.}(2007)\citenamefont
  {Arriz{\'o}n}, \citenamefont {Ruiz}, \citenamefont {Carrada},\ and\
  \citenamefont {Gonz{\'a}lez}}]{Pixelated2007}%
  \BibitemOpen
  \bibfield  {author} {\bibinfo {author} {\bibfnamefont {V.}~\bibnamefont
  {Arriz{\'o}n}}, \bibinfo {author} {\bibfnamefont {U.}~\bibnamefont {Ruiz}},
  \bibinfo {author} {\bibfnamefont {R.}~\bibnamefont {Carrada}},\ and\ \bibinfo
  {author} {\bibfnamefont {L.~A.}\ \bibnamefont {Gonz{\'a}lez}},\ }\bibfield
  {title} {\bibinfo {title} {Pixelated phase computer holograms for the
  accurate encoding of scalar complex fields},\ }\href
  {https://doi.org/https://doi.org/10.1364/JOSAA.24.003500} {\bibfield
  {journal} {\bibinfo  {journal} {JOSA A}\ }\textbf {\bibinfo {volume} {24}},\
  \bibinfo {pages} {3500} (\bibinfo {year} {2007})}\BibitemShut {NoStop}%
\bibitem [{\citenamefont {Ando}\ \emph {et~al.}(2009)\citenamefont {Ando},
  \citenamefont {Ohtake}, \citenamefont {Matsumoto}, \citenamefont {Inoue},\
  and\ \citenamefont {Fukuchi}}]{Mode2009}%
  \BibitemOpen
  \bibfield  {author} {\bibinfo {author} {\bibfnamefont {T.}~\bibnamefont
  {Ando}}, \bibinfo {author} {\bibfnamefont {Y.}~\bibnamefont {Ohtake}},
  \bibinfo {author} {\bibfnamefont {N.}~\bibnamefont {Matsumoto}}, \bibinfo
  {author} {\bibfnamefont {T.}~\bibnamefont {Inoue}},\ and\ \bibinfo {author}
  {\bibfnamefont {N.}~\bibnamefont {Fukuchi}},\ }\bibfield  {title} {\bibinfo
  {title} {Mode purities of laguerre--gaussian beams generated via
  complex-amplitude modulation using phase-only spatial light modulators},\
  }\href {https://doi.org/10.1364/OL.34.000034} {\bibfield  {journal} {\bibinfo
   {journal} {Optics letters}\ }\textbf {\bibinfo {volume} {34}},\ \bibinfo
  {pages} {34} (\bibinfo {year} {2009})}\BibitemShut {NoStop}%
\bibitem [{\citenamefont {Suzuki}\ \emph {et~al.}(2020)\citenamefont {Suzuki},
  \citenamefont {Yang},\ and\ \citenamefont {Hayashi}}]{suzuki2020quantum}%
  \BibitemOpen
  \bibfield  {author} {\bibinfo {author} {\bibfnamefont {J.}~\bibnamefont
  {Suzuki}}, \bibinfo {author} {\bibfnamefont {Y.}~\bibnamefont {Yang}},\ and\
  \bibinfo {author} {\bibfnamefont {M.}~\bibnamefont {Hayashi}},\ }\bibfield
  {title} {\bibinfo {title} {Quantum state estimation with nuisance
  parameters},\ }\href@noop {} {\bibfield  {journal} {\bibinfo  {journal}
  {Journal of Physics A: Mathematical and Theoretical}\ }\textbf {\bibinfo
  {volume} {53}},\ \bibinfo {pages} {453001} (\bibinfo {year}
  {2020})}\BibitemShut {NoStop}%
\bibitem [{\citenamefont {Rosati}\ \emph {et~al.}(2023)\citenamefont {Rosati},
  \citenamefont {Parisi}, \citenamefont {Gianani}, \citenamefont {Barbieri},\
  and\ \citenamefont {Cincotti}}]{rosati2023fundamental}%
  \BibitemOpen
  \bibfield  {author} {\bibinfo {author} {\bibfnamefont {M.}~\bibnamefont
  {Rosati}}, \bibinfo {author} {\bibfnamefont {M.}~\bibnamefont {Parisi}},
  \bibinfo {author} {\bibfnamefont {I.}~\bibnamefont {Gianani}}, \bibinfo
  {author} {\bibfnamefont {M.}~\bibnamefont {Barbieri}},\ and\ \bibinfo
  {author} {\bibfnamefont {G.}~\bibnamefont {Cincotti}},\ }\bibfield  {title}
  {\bibinfo {title} {Fundamental precision limits of fluorescence microscopy: a
  new perspective on minflux},\ }\bibfield  {journal} {\bibinfo  {journal}
  {arXiv preprint arXiv:2306.16158}\ }\href
  {https://doi.org/https://doi.org/10.48550/arXiv.2306.16158}
  {https://doi.org/10.48550/arXiv.2306.16158} (\bibinfo {year}
  {2023})\BibitemShut {NoStop}%
\bibitem [{\citenamefont {Nienhuis}(2017)}]{Analogies2017}%
  \BibitemOpen
  \bibfield  {author} {\bibinfo {author} {\bibfnamefont {G.}~\bibnamefont
  {Nienhuis}},\ }\bibfield  {title} {\bibinfo {title} {Analogies between
  optical and quantum mechanical angular momentum},\ }\href@noop {} {\bibfield
  {journal} {\bibinfo  {journal} {Philosophical Transactions of the Royal
  Society A: Mathematical, Physical and Engineering Sciences}\ }\textbf
  {\bibinfo {volume} {375}},\ \bibinfo {pages} {20150443} (\bibinfo {year}
  {2017})}\BibitemShut {NoStop}%
\bibitem [{\citenamefont {Nienhuis}\ and\ \citenamefont
  {Allen}(1993)}]{Paraxial1993}%
  \BibitemOpen
  \bibfield  {author} {\bibinfo {author} {\bibfnamefont {G.}~\bibnamefont
  {Nienhuis}}\ and\ \bibinfo {author} {\bibfnamefont {L.}~\bibnamefont
  {Allen}},\ }\bibfield  {title} {\bibinfo {title} {Paraxial wave optics and
  harmonic oscillators},\ }\href {https://doi.org/10.1103/PhysRevA.48.656}
  {\bibfield  {journal} {\bibinfo  {journal} {Physical Review A}\ }\textbf
  {\bibinfo {volume} {48}},\ \bibinfo {pages} {656} (\bibinfo {year}
  {1993})}\BibitemShut {NoStop}%
\bibitem [{\citenamefont {Rassias}\ and\ \citenamefont
  {Srivastava}(1992)}]{rassias1992orthogonality}%
  \BibitemOpen
  \bibfield  {author} {\bibinfo {author} {\bibfnamefont {T.~M.}\ \bibnamefont
  {Rassias}}\ and\ \bibinfo {author} {\bibfnamefont {H.}~\bibnamefont
  {Srivastava}},\ }\bibfield  {title} {\bibinfo {title} {The orthogonality
  property of the classical laguerre polynomials},\ }\href
  {https://doi.org/10.1016/0096-3003(92)90124-J} {\bibfield  {journal}
  {\bibinfo  {journal} {Applied mathematics and computation}\ }\textbf
  {\bibinfo {volume} {50}},\ \bibinfo {pages} {167} (\bibinfo {year}
  {1992})}\BibitemShut {NoStop}%
\bibitem [{\citenamefont {Huynh-Thu}\ and\ \citenamefont
  {Ghanbari}(2012)}]{Accuracy2012}%
  \BibitemOpen
  \bibfield  {author} {\bibinfo {author} {\bibfnamefont {Q.}~\bibnamefont
  {Huynh-Thu}}\ and\ \bibinfo {author} {\bibfnamefont {M.}~\bibnamefont
  {Ghanbari}},\ }\bibfield  {title} {\bibinfo {title} {The accuracy of {PSNR}
  in predicting video quality for different video scenes and frame rates},\
  }\href {https://doi.org/10.1007/s11235-010-9351-x} {\bibfield  {journal}
  {\bibinfo  {journal} {Telecommun Syst}\ }\textbf {\bibinfo {volume} {49}},\
  \bibinfo {pages} {35} (\bibinfo {year} {2012})}\BibitemShut {NoStop}%
\bibitem [{\citenamefont {Huynh-Thu}\ and\ \citenamefont
  {Ghanbari}(2008)}]{Scope2008}%
  \BibitemOpen
  \bibfield  {author} {\bibinfo {author} {\bibfnamefont {Q.}~\bibnamefont
  {Huynh-Thu}}\ and\ \bibinfo {author} {\bibfnamefont {M.}~\bibnamefont
  {Ghanbari}},\ }\bibfield  {title} {\bibinfo {title} {Scope of validity of
  psnr in image/video quality assessment},\ }\href
  {https://doi.org/10.1049/el:20080522} {\bibfield  {journal} {\bibinfo
  {journal} {Electronics letters}\ }\textbf {\bibinfo {volume} {44}},\ \bibinfo
  {pages} {800} (\bibinfo {year} {2008})}\BibitemShut {NoStop}%
\end{thebibliography}%

\end{document}